\documentclass[12pt]{article}
\def\macrosPb{}
\usepackage{amsfonts}
\usepackage{amsmath,amssymb,amsthm}
\usepackage{appendix}
\usepackage{bbm} 
\usepackage{amsbsy}
\usepackage{enumerate}
\usepackage{cite}
\usepackage{enumerate}

\def\macrosH{}
\def\macrosHarxiv

\InputIfFileExists{./macros_local.tex}{}{}

\ifdefined\macrosPa
  \usepackage[textwidth=465pt,textheight=650pt,centering]{geometry} 
\else\ifdefined\macrosPb
  \usepackage[textwidth=500pt,textheight=650pt,centering]{geometry} 
\fi\fi

\ifdefined\macrosS


  \usepackage{mathptmx}
  \DeclareMathAlphabet{\mathcal}{OMS}{cmsy}{m}{n}
\fi

\ifdefined\macrosBirk
\else
\usepackage[dvips]{graphicx}
\fi

\ifdefined\macrosSB

\def\UseSection{
        \numberwithin{equation}{section}
	\theoremstyle{plain}
        \newtheorem{theorem}    {Theorem}[section]
        \DefineTheorems 
}

\def\DefineTheorems{
	
	\newtheorem{lemma}      [theorem] {Lemma}
	
	\newtheorem{prop}       [theorem] {Proposition}
	
	\newtheorem{cor}        [theorem] {Corollary}

	\theoremstyle{definition}
	\newtheorem{defn}       [theorem] {Definition}
	
	\newtheorem{example}       [theorem] {Example}

	\theoremstyle{definition}

}

\newcommand{\bt}   {\begin{theorem}}
\newcommand{\et}   {\end  {theorem}}
\newcommand{\bl}   {\begin{lemma}}
\newcommand{\el}   {\end  {lemma}}
\newcommand{\bp}   {\begin{prop}}
\newcommand{\ep}   {\end  {prop}}
\newcommand{\bc}   {\begin{cor}}
\newcommand{\ec}   {\end  {cor}}
\newcommand{\bd}   {\begin{defn}}
\newcommand{\ed}   {\end  {defn}}

\newcommand{\ba}   {\begin{array}}
\newcommand{\ea}   {\end  {array}}
\newcommand{\be}   {\begin{enumerate}}
\newcommand{\ee}   {\end  {enumerate}}
\newcommand{\bi}   {\begin{itemize}}
\newcommand{\ei}   {\end  {itemize}}

\def\eq#1\en{\begin{equation}#1\end{equation}}  
\def\eqsplit#1\ensplit{
	\begin{equation}\begin{split}#1\end{split}\end{equation}
	}
\def\eqalign#1\enalign{
	\begin{align}#1\end{align}
	}
\def\eqmul#1\enmul{
	\begin{multline}#1\end{multline}
	}
\newcommand{\eqarrstar} {\begin{eqnarray*}} 
\newcommand{\enarrstar} {\end{eqnarray*}} 
\newcommand{\eqarray}   {\begin{eqnarray}} 
\newcommand{\enarray}   {\end{eqnarray}} 
\newcommand{\nnb}	{\nonumber \\} 

\newcommand{\lbeq}[1]  {\label{e:#1}}
\newcommand{\refeq}[1] {\eqref{e:#1}}    

\makeatletter
\newcommand{\labelcounter}[2]{{%
	\stepcounter{#1}
	\protected@write\@auxout{}%
	{\string\newlabel{#2}{{\csname the#1\endcsname}{\thepage}}}%
	{\ref{#2}}
	}}
\makeatother
\newcommand{\Nbold} {{\mathbb N}}

\newcommand{\Rbold} {{\mathbb R}}

\newcommand{\Zbold} {{\mathbb Z}}

\newcommand{\Bcal}   {\mathcal{B}} 
\newcommand{\Ccal}   {\mathcal{C}} 
\newcommand{\Dcal}   {\mathcal{D}} 
 
\newcommand{\Fcal}   {\mathcal{F}}

\newcommand{\Kcal}   {\mathcal{K}} 
\newcommand{\Lcal}   {\mathcal{L}} 
 
\newcommand{\Ncal}   {\mathcal{N}} 
 
\newcommand{\Pcal}   {\mathcal{P}}

\newcommand{\Ucal}   {\mathcal{U}} 
\newcommand{\Vcal}   {\mathcal{V}} 
\newcommand{\Wcal}   {\mathcal{W}}

\newcommand{\Zd}    {{ {\Zbold}^d }}

\newcommand{\spose}[1] {{\hbox to 0pt{#1\hss}} }
\newcommand{\ltapprox} {\mathrel{\spose{\lower 3pt\hbox{$\mathchar"218$}}
 \raise 2.0pt\hbox{$\mathchar"13C$}}}
\newcommand{\gtapprox} {\mathrel{\spose{\lower 3pt\hbox{$\mathchar"218$}}
 \raise 2.0pt\hbox{$\mathchar"13E$}}}

\else
\fi

\UseSection   
\usepackage[usenames]{color}

\definecolor{bw}{RGB}{240, 120, 0}
\definecolor{at}{rgb}{0.0, 0.5, 0.0} 

\newcommand{\LT}{{\rm Loc}  }

\newcommand{\DV}{\Dcal}

\renewcommand{\to} {\rightarrow}

\newcommand{\R}{\Rbold}
\newcommand{\Z}{\Zbold}

\newcommand{\N}{\Nbold}
\newcommand{\C}{\mathbb{C}}

\newcommand{\1}{\mathbbm{1}}

\newcommand{\psib}{\bar\psi}

\newcommand{\Ex}{\mathbb{E}}

\newcommand{\chicCov}{{\chi}}

\newcommand{\pair}[1]{\langle #1 \rangle}

\newcommand{\diam}[1]{\textrm{diam}(#1)}

\newcommand{\pt}{{\rm pt}}

\newcommand{\Upt}{U_{\rm pt}}
\newcommand{\Vpt}{V_{\rm pt}}

\newcommand{\h}{\mathfrak{h}}
\ifdefined\macrosSB \else

\fi

\newcommand{\ggen}{\tilde{g}}

\newcommand{\mgen}{\tilde{m}}

\newcommand{\pp}{a}
\newcommand{\qq}{b}

\newcommand{\sigmab}{\bar{\sigma}}

\newcommand{\half}{\textstyle{\frac 12}}

\newcommand{\ddp}[2]{\frac{\partial #1}{\partial #2}}

\newcommand{\epV}{\epsilon_{V}}

\newcommand{\epdV}{\bar{\epsilon}}

\newcommand{\phib}{\bar\phi}

\newcommand{\Kspace}{\Kcal}
\newcommand{\CKspace}{\Ccal\Kspace}

\DeclareMathOperator{\Loc}{Loc} 

\ifdefined\macrosH
  \usepackage{xr-hyper}
  \usepackage{hyperref}
  \hypersetup{hypertexnames=false}

\else\ifdefined\macrosHarxiv
  \usepackage{xr-hyper}
  \usepackage{hyperref}
  \hypersetup{hypertexnames=false, hidelinks}

\else
  \newcommand{\texorpdfstring}[2]{#1}
  \usepackage{xr}
\fi\fi

\hypersetup{
    colorlinks,
    citecolor=black,
    filecolor=black,
    linkcolor=black,
    urlcolor=black
}

\usepackage{mathrsfs}

\newcommand{\csix}{a}
\newcommand{\abar}{{\bar a}}
\newcommand{\ginfty}{s}
\renewcommand{\chicCov}{{\vartheta}}
\newcommand{\thgen}{\tilde{\chicCov}}
\newcommand{\agen}{\tilde{\csix}}

\newcommand{\mg}{\mu}

\newcommand{\onehat}{\hat{1}}

\DeclareMathOperator{\PT}{PT}

\renewcommand{\pp}{{\sf x}}
\renewcommand{\qq}{{\sf y}}

 \title {
   Three-dimensional tricritical spins and polymers
 }
 \date{\vspace{-5ex}} 

 \author{
   Roland Bauerschmidt\thanks{Department of Pure Mathematics and
   Mathematical Statistics, University of Cambridge, Centre for
   Mathematical Sciences, Wilberforce Road, Cambridge, CB3 0WB, UK.
   https://orcid.org/0000-0001-7453-2737.
     E-mail: {\tt rb812@cam.ac.uk}},\;
   Martin Lohmann\thanks{Department of Mathematics,
     University of British Columbia,
     Vancouver, BC, Canada V6T 1Z2.
     Lohmann: https://orcid.org/0000-0002-6627-638X.
     Slade: https://orcid.org/0000-0001-9389-9497.
     E-mail: {\tt marlohmann@math.ubc.ca}, {\tt slade@math.ubc.ca}
     }
     \;
   and Gordon Slade$^\dagger$}

\begin{document}
\maketitle

\begin{abstract}
We consider two intimately related statistical mechanical
problems on $\Z^3$:
(i) the tricritical behaviour of
a model of classical unbounded $n$-component continuous spins with
a triple-well single-spin potential (the $|\varphi|^6$ model),
and (ii) a random walk model
of linear polymers with a three-body repulsion and two-body attraction
at the tricritical \emph{theta point}
(critical point for the collapse transition) where repulsion
and attraction effectively cancel.
The polymer model is exactly equivalent to a supersymmetric spin model which corresponds
to the $n=0$ version of the $|\varphi|^6$ model.
For the spin and polymer models, we identify the tricritical point,
and prove that the tricritical two-point function has Gaussian long-distance
decay, namely $|x|^{-1}$.
The proof is based on an extension of
a rigorous renormalisation group method that has been applied
previously to analyse the $|\varphi|^4$ and weakly self-avoiding walk  models on $\Z^4$.
\end{abstract}



\section{Introduction and main results}

In statistical mechanics, it often occurs that variation of a parameter leads
abruptly to passage from one phase to another as the parameter passes through a critical
value. Prominent examples are freezing, evaporation, superconductivity, Bose-Einstein condensation, or the metal-insulator transition. The mathematically best
understood examples include the Ising model and percolation.
In many cases, the critical point separates an ordered (low-temperature) phase from
a disordered (high-temperature) phase.  The universal behaviour at and near the critical
point is a phenomenon of great interest.

In this paper, we construct the \emph{tricritical point} of
the $n$-component $|\varphi|^6$ model ($n \ge 1$)
on the 3-dimensional cubic lattice $\Z^3$.
The tricritical point is conjecturally
a point of confluence of lines of first-order and second-order phase transition.
We also analyse
a polymer model on $\Z^3$ having a three-body repulsion and two-body attraction, with
parameters adjusted to be at
the tricritical \emph{theta point} where repulsion
and attraction effectively cancel.  For the polymer model, the tricritical point
conjecturally divides a curve of critical points into an arc of self-avoiding walk
critical points and an arc of critical points for polymer collapse.
In each case, we prove Gaussian decay of the tricritical two-point function, i.e. $|x|^{-1}$
decay.

Tricritical behaviour has been much less studied mathematically than critical behaviour.
One reason is that techniques effective in the analysis of the critical behaviour,
such as correlation inequalities, cannot be used to identify a multicritical point.
An exception to this is the renormalisation group (RG) approach.
Our proof is based on the rigorous RG method of \cite{BS-rg-step}, which
has been applied previously to analyse the critical behaviour
of the $n$-component $|\varphi|^4$ model and of the weakly self-avoiding walk,
on the 4-dimensional integer lattice $\Z^4$
(see, e.g., \cite{BBS-phi4-log,BBS-saw4-log,BBS-brief}), as well as for
long-range models below the upper critical dimension \cite{Slad18,LSW17}.
A substantial part of previous papers in this RG scheme---and the main one
that depends on the specific model---is taken up by perturbation theory.
One of the contributions of our work is to simplify
and shorten the treatment of perturbation theory, and a principal focus in the paper
is on this aspect.

The polymer model we study
is exactly equivalent to a supersymmetric spin model which corresponds
to the $n=0$ version of the $|\varphi|^6$ model, and our treatment for $n \ge 1$ and $n=0$
is unified.
The fact that the polymer theta point should be investigated as the
$n=0$ version of the spin problem was clarified in the physics
literature in the 1980s \cite{Dupl86a,Dupl87}.
Although the tricritical theory of $|\varphi|^6$ spins
is a standard part of the physical theory, a mathematical
treatment has been lacking, apart from initial steps taken for a hierarchical model in
\cite{MS86}.  For the polymer problem, there are no previous mathematical results
on the theta point for $\Z^d$;
the combination of repulsion and attraction makes the polymer model even more
difficult than the self-avoiding walk, which is purely repulsive.

We begin with precise definitions of the two models and a precise statement
of our results.

\subsection{$|\varphi|^6$ model}

Fix $L>1$ and
let $\Lambda=\Lambda_N$ denote the 3-dimensional discrete torus of period $L^N$.
We are interested in the infinite-volume limit $N \to \infty$.
We write $\Delta$ for the Laplace operator on functions $f:\Lambda \to \R$,
defined by
\begin{equation} \label{e:Deltadef}
  (\Delta f)_x = \sum_{y \in \Lambda : |y- x|_1=1} (f_y-f_x).
\end{equation}
We use the same symbol $\Delta$ for the Laplacian on $\Z^3$; the meaning should
be clear from context.
The Laplacian operates component-wise on vector-valued functions.
The lattice Green function $C_{0,x}$ is the matrix element of the inverse of $-\Delta_{\Z^3}$,
and its diagonal element $C_{0,0}$ plays a role in our results.

Let $n \ge 1$ be an integer.
The spin field is a function $\varphi : \Lambda \to \R^n$.
Given $a>0$ and $g,\nu \in \R$, let
\begin{equation}\label{e:Vdef}
  V(\varphi_x)
  =
  \frac 18  \csix |\varphi_x|^6
  +
  \frac 14 g |\varphi_x|^4
  +
  \frac 12 \nu |\varphi_x|^2,
\end{equation}
where $|\varphi_x|^2 = \sum_{i=1}^n (\varphi_x^i)^2$.
The \emph{partition function} is
\begin{equation}\label{e:pfcn}
  Z_{\csix,g,\nu;n,N}
  =
  \int_{(\R^n)^\Lambda}
  e^{-\sum_{x \in \Lambda} ( V(\varphi_x) + \frac{1}{2}\varphi_x \cdot (-\Delta \varphi_x))}
  \prod_{x \in \Lambda} d\varphi_x,
\end{equation}
and the expectation of a function $F$ of the spin field is written as
\begin{equation}
    \langle F \rangle_{\csix,g,\nu;n,N}
    =
    \frac{1}{Z_{\csix,g,\nu;n,N}}
    \int_{(\R^n)^\Lambda} F(\varphi)
  e^{-\sum_{x \in \Lambda} ( V(\varphi_x) + \frac{1}{2}\varphi_x \cdot (-\Delta \varphi_x))}
  \prod_{x \in \Lambda} d\varphi_x
  .
\end{equation}
The finite-volume \emph{two-point function} and \emph{susceptibility} are defined
respectively by
\begin{align}
\label{e:2ptn}
  G_{N;0,x}(\csix,g,\nu;n)
  &=
  \frac 1n
  \langle \varphi_0 \cdot \varphi_x \rangle_{\csix,g,\nu;n,N}.
\\
\label{e:suscept}
  \chi_N(\csix,g,\nu;n)
  &=
  \frac 1n \sum_{x\in \Lambda}
  \langle \varphi_0 \cdot \varphi_x \rangle_{\csix,g,\nu;n,N}.
\end{align}
Their $N \to \infty$ limits (assuming they exist) are denoted $G_{0,x}(a,g,\nu;n)$ and
$\chi(\csix,g,\nu;n)$.

\subsection{Polymer model}

The polymer model is defined in terms of
$X=(X(t))_{t \ge 0}$, which denotes the continuous-time simple random walk
on the discrete torus $\Lambda_N$ with nearest-neighbour steps occurring at the events of a
rate-$6$ Poisson process (here ``6" is the degree of $\Lambda_N$).
We write $E_N$ for the expectation when $X(0)=0$.

For $x \in \Lambda$ and $T\ge 0$, the random variable
\begin{equation}
    L_{T,x} = \int_0^T \1_{X(t)=x} \, dt
\end{equation}
denotes the \emph{local time} at $x$ up to time $T$.
For fixed $\csix >0$, and for $g \in \R$ and $x\in \Lambda_N$, we define
\begin{align}
\lbeq{cTN}
    c_{T,N}(\csix,g;x )
    & = E_N
    \left( e^{-\sum_{y\in\Lambda_N} (\csix L_{T,y}^3 + g L_{T,y}^2)}  \1_{X(T)=x}
    \right),
    \\
    c_{T,N}(\csix,g )
    & =
    \sum_{x \in \Lambda_N} c_{T,N}(\csix,g;x )
    = E_N
    \left(   e^{-\sum_{x\in\Lambda_N} (\csix L_{T,x}^3 + g L_{T,x}^2)}
    \right).
\end{align}
Note that, by definition,
\begin{equation}
    \sum_{x \in \Lambda_N}L_{T,x}^2
    = \int_0^T \int_0^T \1_{X(s)=X(t)} \, ds \, dt,
    \qquad
    \sum_{x \in \Lambda_N}L_{T,x}^3
    = \int_0^T \int_0^T \int_0^T \1_{X(s)=X(t)=X(u)} \, ds \, dt  \, du.
\end{equation}
We are interested in $g<0$; in this case there is an attractive two-body
term and a competing repulsive three-body term in the exponent on the right-hand side of
\refeq{cTN}.

The finite-volume \emph{two-point function} and
\emph{susceptibility} are defined respectively by
\begin{equation}
    G_{N;0,x}(a,g,\nu;0) = \int_0^\infty c_T(\csix,g;x) e^{-\nu T} \, dT,
    \qquad
    \chi_N(\csix,g,\nu;0) = \int_0^\infty c_T(\csix,g) e^{-\nu T} \, dT.
\end{equation}
The finite-volume susceptibility is finite for all
$(g,\nu)\in \R^2$ (and hence so is the two-point function),
since by H\"older's inequality
$T = \sum_{x\in \Lambda} L_{T,x}
\le (\sum_{x\in\Lambda}L_{T,x}^3)^{1/3}|\Lambda|^{2/3}$, and also
$\sum_{x\in\Lambda}L_{T,x}^2 \le (\sup_x L_{T,x})\sum_x L_{T,x} \le T^2$, so
\begin{equation}
    c_{T,N}(a,g)e^{-\nu T} \le e^{-aT^3|\Lambda|^{-2} + |g| T^2 +|\nu|T} .
\end{equation}
The right-hand side is indeed integrable for all $(g,\nu)\in \R^2$, as long as $a>0$.

We define the infinite-volume two-point function and susceptibility by
\begin{equation}
\lbeq{wsaw-suscept}
    G_{0,x}(a,g,\nu;0) = \lim_{N\to\infty}G_{N;0,x}(a,g,\nu;0),
    \qquad
    \chi_(\csix,g,\nu;0) = \lim_{N\to\infty}\chi_N(\csix,g,\nu;0),
\end{equation}
assuming the limits exist.

\subsection{Supersymmetry and $n=0$}

The two-point function for the polymer model can be written
exactly as the supersymmetric integral
\begin{equation}
\label{e:intrep}
  G_{N;0,x}(\csix,g,\nu;0)
  =
  \int
  e^{-\sum_{y\in\Lambda_N} (\csix \tau_y^3 + g \tau_y^2 + \nu\tau_y )}
  e^{-S}
  \phib_0\phi_x,
\end{equation}
with
\begin{align}
    \tau_y &= \phi_y\phib_y + \psi_y \psib_y,
    \quad
    S = \sum_{y\in\Lambda_N} \left( \phi_y(-\Delta\phib)_y + \psi_y (-\Delta\psib)_y \right).
\end{align}
The above notation is explained in \cite[Chapter~11]{BBS-brief} and the identity
\refeq{intrep} is an immediate consequence of \cite[Corollary~11.3.7]{BBS-brief}.
The analysis of the supersymmetric model is a modification of the analysis of
the $|\varphi|^6$ model, which follows the same well-trodden path as for the 4-dimensional
analysis in \cite{ST-phi4,BBS-saw4}.  Formulas for the spin system involving
the number $n$ of spin components transfer to the polymer setting with $n=0$.
For notational simplicity, we focus
our discuss in this
paper on the case $n \ge 1$, and comment occasionally on the supersymmetric case.

\subsection{Main result}

Our main result is Theorem~\ref{thm:main-n}.  The existence of
the limit defining the tricritical two-point function,
namely the left-hand side of \refeq{G-asy}, is part of its statement.
For $n \ge 1$ the point
$(g_c(a),\nu_c(a))$
is the tricritical point, and for $n=0$ it is
the tricritical theta point.
In terms of critical exponents, \refeq{G-asy} says that
tricritical two-point function has decay $|x|^{-(d-2+\eta)}$ with $\eta =0$.

\begin{theorem}
\label{thm:main-n}
Let $d=3$ and $n \ge 0$.
Let $L$ be sufficiently large and $\delta>0$ sufficiently small.
There exists a continuous function
$(g^*(t,a),\nu^*(t,a))$ of $(t,a)\in (0,\delta)^2$, with limit
$(g_c(a),\nu_c(a))=\lim_{t\downarrow 0}(g^*(t,a),\nu^*(t,a))$, such that
for $(t,a)\in (0,\delta)^2$
the limit
\begin{equation}
    G_{0,x}(a,g^*(t,a),\nu^*(t,a);n)
    = \lim_{N\to\infty}G_{N;0,x}(a,g^*(t,a),\nu^*(t,a);n)
\end{equation}
 exists,
and moreover
\begin{alignat}{2}
\lbeq{G-asy}
    \lim_{t \downarrow 0}
    G_{0,x}(a,g^*(t,a),\nu^*(t,a);n)
    & =
    A_{a,n}\frac{1}{|x|} \left( 1+O((\log|x|)^{-1}) \right)
\end{alignat}
as $x\to\infty$, with $A_{a,n}=(4\pi)^{-1}(1+O(a))$.
\end{theorem}

It would be of interest to study the
geometry of the curve $(g^*(t,a),\nu^*(t,a))$ as $t$ varies,
but this has not yet been investigated.
It would also be of interest to consider
the approach to the tricritical point from a more general direction;
see Section~\ref{sec:phase-diagram}.

\begin{theorem}
\label{thm:tricrit-pt}
For $n \ge 0$, the asymptotic behaviour of the tricritical point
$(\nu_c(a),g_c(a))$, as $a \downarrow 0$, is given
(with $C_{0,0}=(-\Delta_{\Z^3})^{-1}_{0,0}$) by
\begin{align}
    g_c(a) & = -\frac{3}{2} (n+4)C_{0,0} \csix + O(a^2),
    \\
    \nu_c(a) & = \frac{3}{4}(n+4)(n+2) C_{0,0}^2 + O(a^2) .
\end{align}
\end{theorem}

\subsection{Discussion}

\subsubsection{Conjectured phase diagram}
\label{sec:phase-diagram}

\begin{figure}[ht]
\begin{center}
\input{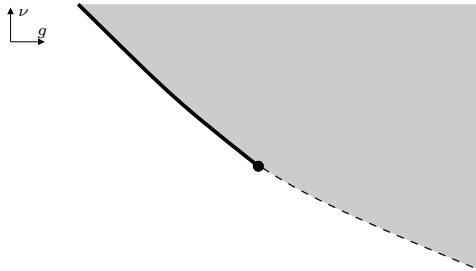}
\end{center}
\caption{
The conjectured phase diagram for $n \ge 0$.
For $n \ge 1$, the tricritical point lies at the confluence of lines of first-order (solid)
and second-order (dashed) phase transitions.  For $n=0$, the dashed line corresponds to
self-avoiding walk critical points, while the solid line corresponds to polymer collapse.
The shaded region is the disordered phase.}
\label{fig:tricritical-phase-diagram}
\end{figure}

The conjectured phase diagram associated with the tricritical point,
as predicted by the Landau mean-field theory
(see, e.g., \cite[Section~7.6.4]{Arov18} or \cite[Appendix~5.A]{ID89b}),
is illustrated in Figure~\ref{fig:tricritical-phase-diagram}.

A complete analysis of the phase diagram for the polymer model on the complete
graph (mean-field theory) is given in  \cite{BS20}.
We expect that the phase diagram on $\Z^d$ is qualitatively
the same for all $d\geq 3$.

Theorem~\ref{thm:main-n} indicates that Gaussian decay of the two-point function
occurs at the point $(g_c(a),\nu_c(a))$, but does not show that it is a tricritical point
in the sense of being a point of confluence of first- and second-order transitions.
However, for dimension $d=3$, we expect that the tricritical point is the only
point in the $(g,\nu)$ plane where Gaussian decay occurs.
A more complete description would require an analysis of the neighbourhood
of $(g_c,\nu_c)$, a very difficult problem.

There is potential to extend our methods to study the divergence of the
susceptibility (for $n \ge 0$) and specific heat (for $n \ge 1$)
as the tricritical point is approached from the disordered phase.
Our preliminary calculations support the conjecture that the open half plane
\begin{equation}
    H_n = \{ (g,\nu)\in \R^2 : \nu + (n+2)C_{0,0} g > 0\}
\end{equation}
plays a key role for this.  In particular, given a point $(g,\nu)$ such that
$(g-g_c,\nu-\nu_c) \in H_n$, we conjecture that the susceptibility and specific
heat are finite along the line segment $(g_c+t(g-g_c),\nu_c+t(\nu-\nu_c))$ for $t\in (0,t_*]$
(for $t_* $ small  depending on $a$),
that the susceptibility asymptotically
diverges as a multiple of $t^{-1}$ as $t \downarrow 0$, and
that the specific heat asymptotically diverges as a multiple of $t^{-1/2}$.
The conjecture assumes that $a$ is taken smaller as the angle of approach
becomes closer to the boundary of $H_n$.

Our present methods are not sufficient to study the high-density phase.
The related problem of
the high-density phase of the weakly self-avoiding walk
on a $4$-dimensional hierarchical lattice has been studied in \cite{GI95}.
Recent progress on applying the RG to study broken symmetry
was obtained in \cite{Lohm19}, where the critical magnetisation of the 4-dimensional
$\varphi^4$ model is analysed.

\subsubsection{Infrared asymptotic freedom}

For dimension $d=4$, the mean-field $|x|^{-2}$ decay of the critical two-point
function has
been extensively studied \cite{GK85,BBS-saw4,ST-phi4,FMRS87}, both for spin
and polymer (weakly self-avoiding walk) models.  On $\Z^4$, the mean-field
behaviour results from infrared asymptotic freedom, which is itself closely
connected to the marginal nature (in the RG sense) of $|\varphi|^4$ for $d=4$.
The RG flow is to a Gaussian fixed point.
The logarithmic divergence of the massive bubble diagram $\sum_{x\in \Z^4}
[(-\Delta+m^2)_{0,x}^{-1}]^2$ as $m^2 \downarrow 0$ plays an important role,
especially in the study of the divergence of thermodynamic quantities in
the approach to the critical point where logarithmic corrections
appear \cite{BBS-saw4-log,BBS-phi4-log,HT87}.

For dimension $d=3$, $|\varphi|^4$ becomes a relevant monomial, whereas $|\varphi|^6$
is marginal.
The $|x|^{-1}$ decay in \refeq{G-asy} is again mean-field
behaviour, which is again a consequence of infrared asymptotic freedom, and
correspondingly a Gaussian RG fixed point.
The role of the bubble diagram is now played instead
by the logarithmically divergent diagram $\sum_{x\in \Z^4} [(-\Delta+m^2)_{0,x}^{-1}]^3$;
the bubble diagram diverges
as $m^{-1}$.

While both are governed by the Gaussian free field fixed point,
a difference between the tricritical theory for $\Z^3$ and the critical theory for $\Z^4$ is that for the latter it is only required
to tune one parameter (coefficient of $|\varphi|^2$)
to obtain a critical theory, whereas for $\Z^3$ it is necessary
to tune two parameters (coefficients of $|\varphi|^2$ and $|\varphi|^4$) to obtain
the tricritical theory.
This has long been understood in the physics literature, e.g.,  \cite{SAS75,WR73}.

Much of our proof of Theorem~\ref{thm:main-n} parallels the analysis used for
the 4-dimensional case in \cite{BBS-saw4,ST-phi4}.  In particular, with minor modifications,
the results in \cite{BS-rg-IE,BS-rg-step} provide the analysis of a single RG step.
The tuning of parameters follows as in \cite{BBS-rg-flow} with only notational changes.
The analysis of perturbation theory is also similar to that in \cite{BBS-saw4,ST-phi4},
but here we present an improved and streamlined treatment of perturbation theory, which
in particular demystifies the change of variables used in \cite{BBS-rg-pt}.

\subsubsection{Interacting self-avoiding walk}

The tricritical model of the theta point studied in this paper was investigated
extensively in the physics literature during the 1980s, e.g., \cite{Dupl86a,Dupl87}
where misconceptions in the earlier literature were clarified.
We are not aware of any previous mathematically rigorous analysis.
In the mathematics literature, it has been more common to study the \emph{interacting
self-avoiding walk}, in which a self-avoiding or weakly self-avoiding walk receives
an energetic reward for nearest-neighbour contacts \cite{Holl09}.
The interacting self-avoiding
walk has been studied in dimensions $d>4$ in \cite{Uelt02,HH19}, in dimension $d=4$
in \cite{BSW-saw-sa}, and in dimension $d=1$ in \cite{HK01,HKK02}.
The interacting
prudent walk was studied in \cite{PT18}.
Recent numerical work for $d=2$ appears in \cite{BGJ20}.

\section{The RG map and RG flow}

Our RG method is based on a multiscale analysis which is implemented via the
finite-range covariance decomposition discussed in Section~\ref{sec:frd}.
Representations for the susceptibility and the two-point function are presented
in Section~\ref{sec:chiG}.  In Section~\ref{sec:RGmap}, the RG map is discussed
along with the estimates on the global RG flow which are
an essential ingredient for the proof of our main result.   The tricritical point
is identified in Section~\ref{sec:tript}.

\subsection{Covariance decomposition and progressive integration}
\label{sec:frd}

We generalise \eqref{e:Vdef} and define
\begin{equation}
  V_{\csix,g,\nu,z}(\varphi_x)
  =
  \tfrac{1}{8}\csix |\varphi_x|^6 +
  \tfrac{1}{4} g |\varphi_x|^4 + \half\nu|\varphi_x|^2
  + \half z \varphi_x(-\Delta\varphi)_x .
\end{equation}
By definition, for any \emph{mass} $m > 0$ and for any
\emph{wave function renormalisation} $z_0 > -1$,
\begin{equation} \label{e:Vsplit}
  V_{\csix,g,\nu,1}(\varphi_x)
  = V_{0,0,m^2,1}((1+z_0)^{-1/2}\varphi_x)
  + V_{\csix_0,g_0,\nu_0,z_0}((1+z_0)^{-1/2}\varphi_x)
  ,
\end{equation}
with
\begin{equation} \label{e:g0gnu0nu}
  \csix_0 = \csix (1+z_0)^3, \quad g_0 = g(1+z_0)^2, \quad \nu_0 = (1+z_0)\nu-m^2
  .
\end{equation}
For $X \subset \Lambda$, we define
\begin{equation}
\lbeq{V0Z0}
    V_0(X) = V_0(\varphi,X) = \sum_{x\in X} V_{\csix_0,g_0,\nu_0,z_0}(\varphi_x),
    \qquad
    Z_0(\varphi) = e^{-V_{0}(\varphi,\Lambda_N)}  .
\end{equation}

Given a {positive semi-definite covariance $\Lambda \times \Lambda$ matrix $C$,
we write expectation with respect
to the Gaussian measure for $n$-component fields as $\Ex_C$.
For $n=0$, we instead use a superexpectation, exactly as in \cite{BBS-saw4-log}.
Let $C=(-\Delta_{\Lambda_N}+m^2)^{-1}$, where $\Lambda_N$ is
the discrete torus of period $L^N$.  For $m^2>0$, the inverse matrix exists and it is
positive definite.
Then, with $\tilde F (\varphi) = F((1+z_0)^{1/2}\varphi)$,
the change of variables $\varphi_x \mapsto (1+z_0)^{1/2}\varphi_x$ gives
\begin{equation}
\lbeq{ExF}
    \pair{F}_{\csix,g,\nu,N}
    = \frac{\Ex_C \tilde F Z_0}{\Ex_C Z_0} .
\end{equation}

We use decompositions of both of the covariances
$(-\Delta_\Zd + m^2)^{-1}$ and $(-\Delta_{\Lambda_N} +m^2)^{-1}$, based
on the method of \cite{Baue13a}.
For $\Zd$, this Green function exists for $d>2$ for all
$m^2 \ge 0$, but for finite $\Lambda$ we must restrict to $m^2>0$.
As we discuss in Appendix~\ref{app:decomp},
there is a sequence
$(C_j)_{1 \le j < \infty}$ (depending on $m^2 \ge 0$)
of positive-definite 
covariances $C_j$ on $\Zd$ such that
\begin{equation}
\lbeq{ZdCj}
    (\Delta_\Zd + m^2)^{-1} = \sum_{j=1}^\infty C_j
    \quad
    \quad
    (m^2 \ge 0).
\end{equation}
The $C_j$ are translation invariant and have the \emph{finite-range} property
$C_{j;x,y} = 0$ if $|x-y| \geq \frac{1}{2} L^j$.
Thus, for $j<N$, $C_j$ can also be identified as a
covariance on the torus $\Lambda$.
For $m^2>0$, there is also a covariance $C_{N,N}$ on $\Lambda$ such that
\begin{equation}
\lbeq{NCj}
    (-\Delta_\Lambda + m^2)^{-1} = \sum_{j=1}^{N-1} C_j + C_{N,N}
    \quad
    \quad
    (m^2 > 0)
    ,
\end{equation}
so the finite-volume and infinite-volume decompositions agree
until the last term in the finite-volume decomposition.
Properties of the covariance decomposition are collected in Appendix~\ref{app:decomp}.

We define the \emph{mass scale} $j_m$ to be the largest integer $j$ such that $mL^{j} \le 1$.
In particular, $\lim_{m^2 \downarrow 0}j_m=\infty$.
By Proposition~\ref{prop:Cdecomp},
for multi-indices $\alpha,\beta$ with
$\ell^1$ norms $|\alpha|_1,|\beta|_1$ at most some fixed value $p$,
\begin{equation}
  \label{e:scaling-estimate-theta}
  |\nabla_x^\alpha \nabla_y^\beta C_{j;x,y}|
  \leq c \chicCov_{j-1}
  L^{-(j-1)(2[\varphi]+(|\alpha|_1+|\beta|_1))}
  ,
\end{equation}
where
\begin{equation}
\lbeq{chicCovdef}
    \chicCov_j = 2^{-(j-j_m)_+}
\end{equation}
with $x_+=\max\{x,0\}$, where the \emph{dimension} of the field is
\begin{equation}
\lbeq{dimphi}
    [\varphi]=\frac {d-2}{2} = \frac 12,
\end{equation}
and where the constant $c$ depends on $p$ but not on $m^2,j,L$.
The bound \eqref{e:scaling-estimate-theta} also holds for $C_{N,N}$ if
$m^2L^{2(N-1)} \ge \delta$ for some fixed $\delta >0$,
with $c$ depending on $\delta$ but not on $N$.

For $n \ge 1$,
we write $\Ex_C\theta F$  for the convolution of $F$ with the Gaussian expectation $\Ex_C$.
Explicitly,
$\theta$ is the shift operator $\theta F(\varphi,\zeta)=F(\varphi + \zeta)$, and
\begin{equation}
\lbeq{thetadef}
  (\Ex_C\theta F)(\varphi) = \Ex_C F(\varphi+\zeta).
\end{equation}
In \refeq{thetadef}, the expectation $\Ex_C$ acts on $\zeta$ with $\varphi$ held fixed.
By a standard property of Gaussian integration,
the decomposition \refeq{NCj} gives
\begin{equation}
    \label{e:progressive}
    \Ex_{C}\theta F
    =
    \big( \Ex_{C_{N,N}}\theta \circ \Ex_{C_{N-1}}\theta \circ \cdots
    \circ \Ex_{C_{1}}\theta\big) F
    .
\end{equation}
With $Z_0 = e^{-V_0(\Lambda_N)}$ as in \refeq{V0Z0},
we define
\begin{equation}
\label{e:Zjdef}
  Z_{j+1} = \Ex_{C_{j+1}}\theta Z_j \quad\quad
  (j<N).
\end{equation}
In particular,
\begin{equation}
\label{e:ZN}
Z_N = \Ex_C\theta Z_0,
\qquad
Z_N(0)=\Ex_CZ_0.
\end{equation}
To simplify the notation, we write $\Ex_{j} = \Ex_{C_j}$,
and leave implicit the dependence of $C_j$
on $m^2$.  There is a supersymmetric version of \refeq{progressive}--\refeq{ZN},
exactly as in \cite{BBS-saw4-log}.

\subsection{Susceptibility and two-point function}
\label{sec:chiG}

\subsubsection{Susceptibility}

Let $n \ge 1$.
We define
\begin{equation}
\lbeq{chihatdef}
    \hat\chi_N(m^2,\csix_0,g_0,\nu_0,z_0)
    =
    \sum_{x\in \Lambda_N} \frac{\Ex_C(\zeta_0^{1} \zeta_x^{1} Z_0(\zeta))}{\Ex_C ( Z_0(\zeta))}.
\end{equation}
By \refeq{ExF}, when \refeq{g0gnu0nu}
is satisfied the finite volume susceptibility obeys
\begin{equation}
\lbeq{chichihat}
    \chi_N(\csix,g,\nu) = (1+z_0) \hat\chi_N(m^2,\csix_0,g_0,\nu_0,z_0).
\end{equation}
Although \refeq{chichihat} requires \refeq{g0gnu0nu}, it is nevertheless
useful at times to relinquish the identity
and consider the variables $(m^2,\csix_0,g_0,\nu_0,z_0)$
on the right-hand side of \refeq{ExF} as independent variables.

Given a test function $J : \Lambda_N \to \R^n$, let
\begin{equation}
\lbeq{Sigdef}
  \Sigma_N(J) = \Ex_{C}(Z_0(\zeta) e^{(\zeta, J)})
  ,\qquad \text{where $(\zeta, J) = \sum_{x\in \Lambda_N}\sum_{i=1}^n \zeta_x^iJ_x^i$}
  .
\end{equation}
Differentiation in the direction $\1=(1,0,\dots,0)$ gives
\begin{equation}
\lbeq{chiSig}
    \hat\chi_N(m^2,\csix_0,g_0,\nu_0,z_0)
=
    \frac{1}{|\Lambda_N|} \frac{D^2 \Sigma_N(0;\1,\1)}{Z_N(0)}.
\end{equation}
By completing the square, we obtain
\begin{equation}
\lbeq{LTC}
  \Sigma_N(J)
  = e^{\frac 12 (J ,CJ)}
  \Ex_{C}(Z_0(\zeta + CJ))
  = e^{\frac 12 (J ,CJ)}
  Z_N(CJ).
\end{equation}
Differentiation of \refeq{LTC}, together with $C\1 = m^{-2}\1$, leads to
\begin{equation}
    D^2 \Sigma_N(0;\1,\1)
    = \frac{1}{m^2} |\Lambda_N| Z_N(0) + \frac{1}{m^4} D^2 Z_N(0;\1,\1),
\end{equation}
and hence
\begin{equation}
\lbeq{chiZN}
    \hat\chi_N(m^2,\csix_0,g_0,\nu_0,z_0) =
    \frac{1}{m^2}
    +
    \frac{1}{m^4} \frac{1}{|\Lambda_N|} \frac{D^2 Z_N(0;\1,\1)}{Z_N(0)}.
\end{equation}
Thus the evaluation of the susceptibility reduces to the evaluation of $Z_N(\varphi)$
for a constant field $\varphi$.

\subsubsection{Two-point function and observable fields}
\label{sec:of}

Let $n \ge 1$.
By \refeq{ExF}, when \refeq{g0gnu0nu} is satisfied
the two-point function \refeq{2ptn} can be written as
\begin{equation}
  G_{N;\pp,\qq}(a,g,\nu;n)
  = (1+z_0) \frac{\Ex_C (\zeta_\pp^1\zeta_\qq^1 Z_0(\zeta))}{\Ex_C Z_0(\zeta)}
  .
\end{equation}
Let $\onehat = n^{-1/2}(1,\ldots,1)\in \R^n$.
Given $\pp,\qq\in\Z^d$ and given \emph{observable fields}
$\sigma_\pp,\sigma_\qq \in \R$, we define $V_0$  by
\begin{equation}
\label{e:V0n1}
    V_{0;x}
    =
    V_{0;x}^\varnothing  - \sigma_\pp (\onehat \cdot \varphi_\pp)   \1_{x=\pp}
    - \sigma_\qq (\onehat \cdot \varphi_\qq ) \1_{x=\qq},
\end{equation}
with $V_0^\varnothing$ a new notation for the bulk polynomial \refeq{V0Z0}.
Although $\sigma_\pp$ and $\sigma_\qq$ carry subscripts $\pp,\qq$,
they are real constants.
Let $\partial_{\sigma_\pp\sigma_\qq}^2$ denote $\ddp{^2}{\sigma_\pp\sigma_\qq}|_{\sigma_\pp = \sigma_\qq = 0}$.
Then
\begin{equation}
\label{e:corrdiff}
    G_{N;\pp,\qq}(a,g,\nu;n)
    =
    (1+z_0)
    \partial_{\sigma_\qq\sigma_\qq}^2 \log \Ex_C e^{- V_0 
    (\Lambda)}.
\end{equation}

The observable field
$\sigma_\pp \onehat \1_{x=\pp} + \sigma_\qq \onehat\1_{x=\qq}$
can be regarded as an implementation of a special
case of the test function $J$ used in \refeq{Sigdef}.  However, for the susceptibility
only a constant test function was needed, whereas the observable fields are
highly localised.  For the two-point function, we now regard the
observable field as part of the potential $V$, and we track the flow of the new terms
in $V$ under progressive integration.  It is via this flow that we will be able to
compute the behaviour of the two-point function.

A hybrid approach is also possible, as follows.  Let
$\partial_{\sigma_\pp}$ denote $\ddp{}{\sigma_\pp}|_{\sigma_\pp = \sigma_\qq = 0}$
and $Z^\varnothing_N(0) = \Ex_C e^{- V_0^\varnothing(\Lambda)}$.
Let $\hat\1$ denote the constant field $\hat\1_x = \onehat$ for all $x \in \Lambda$.
We note for later use that, with $Z_0$ now defined using \refeq{V0n1},
\begin{align}
\lbeq{lam1identity}
    \hat\chi_N
    &= \frac{\partial_{\sigma_\pp} D \Sigma_N(0;\hat\1)}{Z_N^{\varnothing}(0)}
    =  \frac{\partial_{\sigma_\pp} D Z_N(0;C\hat\1)}{Z_N^{\varnothing}(0)}
    =  \frac{1}{m^2}
    \frac{\partial_{\sigma_\pp} D Z_N(0;\hat\1)}{Z_N^{\varnothing}(0)}
    ,
\end{align}
where
the second equality follows from \refeq{LTC} and $C\1 = m^{-2}\1$.

The calculation of the two-point function only requires the second derivative in
\refeq{corrdiff}, and therefore only needs the dependence on the observables
to second order.
As in \cite{BBS-saw4,ST-phi4,BS-rg-step}, we formalise this simplification via use of
a quotient space,
in which two functions of $\varphi,\sigma_\pp,\sigma_\qq$ become equivalent if
their formal power series in the observable fields agree to
order $1,\sigma_\pp, \sigma_\qq, \sigma_\pp \sigma_\qq$.
Thus we define $\Ncal$ to consist of $C^{16}$
functions of $\varphi$ of the form
\begin{equation}
\label{e:Fdecomp}
F = F^\varnothing + \sigma_\pp  F^\pp + \sigma_\qq  F^\qq  +\sigma_\pp \sigma_\qq F^{\pp\qq},
\end{equation}
where each $F^\alpha$ is a function of $\varphi$.

For $n =0$, the above can be modified exactly as
in \cite{BBS-saw4},
with \emph{observable fields} $\sigma_\pp,\sigma_\qq \in \mathbb{C}$, and
\begin{equation}
\label{e:V0n0}
V_{0;x}  =
V^\varnothing_{0;x}   - \sigma_\pp \phib_{\pp} \1_{x=\pp} - \sigmab_\qq \phi_{\qq} \1_{x=\qq}.
\end{equation}
Similarly to \eqref{e:corrdiff},
since the partition function with $\sigma_\pp =\sigmab_\qq=0$ is equal to $1$
by supersymmetry, with $\Ex_C$ now a super-expectation
we have
\begin{equation}
\label{e:intobs}
G_{N;\pp,\qq}(a,g,\nu; 0)
= (1+z_0)
\partial_{\sigma_\pp\sigmab_\qq}^2
\Ex_C e^{- V_{0} (\Lambda)}
= (1+z_0)
\partial_{\sigma_\pp \sigmab_\qq}^2
\log \Ex_C e^{- V_{0} (\Lambda)}
.
\end{equation}

\subsection{RG map}
\label{sec:RGmap}

Integration of a single scale is recorded in \refeq{Zjdef} as $Z_{j+1}=\Ex_{C_{j+1}}\theta Z_j$,
with $Z_0 = e^{-V_0(\Lambda)}$.  Now we use the version \refeq{V0n1} of $V_0$ with
observables.  It would be desirable to have $Z_j$ also
represented by an \emph{effective potential} $V_j$ as $Z_j \approx e^{-V_j(\Lambda)}$
with $V_j$ of the same form as $V_0$ but with renormalised coupling constants.
However, such an approximation requires great care.  Instead, exactly as in
\cite{BS-rg-step}, we use a representation involving the circle product.
In the following, we do not always give precise definitions, as these can be found
in \cite[Section~1]{BS-rg-step} with the same notation as used here.

We need the following definitions.
For each $j=0,1,\ldots,N$, the discrete torus $\Lambda_N$ of period $L^N$
partitions into
$L^{N-j}$ disjoint $d$-dimensional cubes of side $L^j$, called {\em blocks}, or $j$-{\em blocks}.
We denote the set of $j$-blocks by ${\cal B}_j$.
A union of $j$-blocks (possibly empty) is called a {\em polymer} or $j$-\emph{polymer},
and the set of $j$-polymers is denoted ${\cal P}_j$.
The set of blocks in a polymer $X \in \Pcal_j$ is denoted $\Bcal_j(X)$,
and the set of connected components $\Ccal_j(X)$.
The unique $N$-block is $\Lambda_N$ itself.
With these definitions,
we write each $Z_j$ in the form
\begin{gather}\label{e:ZIK}
    Z_j =  e^{\xi_j}(I_j \circ K_j)(\Lambda)
    =
    e^{\xi_j} \sum_{X \in \Pcal_j} I_j(\Lambda \setminus X)K_j(X),
\end{gather}
where $I_j(X)$ and $K_j(X)$ are functions of the field in the neighbourhood of $X$,
and the prefactor has the form
\begin{align}
     \xi_j & = -u_j|\Lambda| + \frac 12( q_{\pp,j}+q_{\qq,j})\sigma_\pp\sigma_\qq.
\end{align}
The functions $I_j$ and $K_j$ satisfy the factorisation properties
$I_j(X) = \prod_{B\in\Bcal_j(X)} I_j(B)$  and $K_j(X) = \prod_{C \in \Ccal_j(X)} K_j(C)$.
More precisely,}
$I_j(B)=e^{-V_j(B)}(1+W_j(V_j,B))$, where $W_j(V_j,B)$ is an explicit quadratic function
(defined in \refeq{WLTF} below) of
\begin{align}
\lbeq{Vjdef}
    V_{j,x}  & =  a_j \tau_x^3 +
     g_j \tau_x^2 +  \nu_j \tau_x + z_j \tau_{\Delta,x}
     -\lambda_{\pp,j}\sigma_\pp (\onehat \cdot \varphi_x)\1_{x=\pp}
     -\lambda_{\qq,j}\sigma_\qq (\onehat\cdot \varphi_x)\1_{x=\qq}
     .
\end{align}
The nonperturbative coordinate $K_j$ encompasses all irrelevant (in RG sense) terms
that are $O(V_j^3)$.  All first- and second-order contributions are in $I_j$.
Since $W_j(V_j)=O(V_j^2)$,
it is instructive to pretend that $I_j(X) \approx e^{-V_j(X)}$.
At the last scale, \eqref{e:ZIK}
becomes a sum over the two polymers $X=\varnothing$ and $X=\Lambda_N$, and hence
\begin{equation}
\lbeq{ZN2terms}
  Z_N
  = e^{\xi_N}(I_N(\Lambda_N) + K_N(\Lambda_N))
  = e^{\xi_N}[e^{-V_N(\Lambda_N)}(1+W_N(\Lambda_N)) + K_N(\Lambda_N)]
  .
\end{equation}

If the above can be achieved with $K_N$ appropriately vanishing in the limit $N \to \infty$,
then we would have $Z_N(\varphi) \approx e^{\xi_N - V_N(\varphi)}$, and differentiation
as in \refeq{chiZN} and \refeq{corrdiff} would permit the susceptibility and two-point function
to be evaluated as
\begin{equation}
    \chi_N \approx (1+z_0)(m^{-2}+m^{-4} \nu_N),
    \qquad
    G_{N;\pp,\qq} \approx (1+z_0) q_N.
\end{equation}
This suggests that knowledge of the coupling constants $\nu_N$ and $q_N$ is tantamount
to a computation of the susceptibility and two-point function.

The calculation of the coupling constants
requires an understanding of iterations of the \emph{RG map},
which is a map
\begin{equation}
\label{e:RGmap}
     ( V_j, K_j) \mapsto ( U_{j+1},K_{j+1}) = (\delta \xi_{j+1},V_{j+1},K_{j+1})
\end{equation}
that is defined in such a way that
\begin{equation}
    \label{e:Kspace-objective}
    Z_{j+1}
    =
    \Ex_{j+1}\theta Z_j
    =
    e^{\xi_j}
    \Ex_{j+1}\theta (I_j \circ K_j)(\Lambda)
    =
    e^{\xi_{j+1}}(I_{j+1} \circ K_{j+1})(\Lambda)
    ,
\end{equation}
with
\begin{equation}
    \delta\xi_{j+1}=\xi_{j+1}-\xi_j
    =
    -\delta u_{j+1}|\Lambda| +
    \frac 12(\delta q_{\pp,j}+ \delta q_{\qq,j})\sigma_\pp\sigma_\qq.
\end{equation}
Indefinite iteration of the RG map requires tuning of the initial values $g_0,\nu_0$
to values corresponding to the tricritical point.

The map $(V_j,K_j) \mapsto (\delta\xi_{j+1},V_{j+1})$ is a combination of a perturbative map $\PT$,
which is a function only of $V_j$,
and a nonperturbative remainder which extracts the relevant part of $K_j$.
Thus,
\begin{align}
    V_{j+1} &= \PT_j^{(0)}(V_j) + R_{j+1}^{(0)}(V_j,K_j),
    \\
    \delta\xi_{j+1} & = \PT_j^{(\xi)}(V_j) + R_{j+1}^{(\xi)}(V_j,K_j).
\end{align}
The maps $\PT_j^{(0)}, \PT_j^{(\xi)}$
are components of an explicit quadratic map $\PT_j$ defined in Section~\ref{sec:PTdef}.
[In detail, $\PT^{(0)}$ is the operator $\PT$ followed by the replacement
of $(y,z,u,q_\pp,q_\qq)$ by $(0,y+z,0,0,0)$,
and $\PT^{(\xi)}$ consists of the $(u,q_\pp,q_\qq)$
components of $\PT$.]
We write the individual coefficients of the
remainder terms as $R_{j+1}^a, R_{j+1}^g, \ldots$.

The next theorem is for the bulk, which corresponds to setting $\sigma_\pp=\sigma_\qq =0$.
It asserts the existence of a global bulk RG flow with estimates on $K_j$ and on the remainders
$R_j$ to perturbation theory.
The statement of the theorem involves seminorms $\|F\|_{T_0(\ell_j)}$ (defined
in \cite[Section~2.3]{BBS-phi4-log} for $n \ge 1$ and in \cite[Section~6.3]{BBS-saw4-log}
for $n=0$).
Although the proof of the theorem requires working with a stronger norm
(see Section~\ref{sec:normparameters}),
for our application we only need to know that
\begin{equation}
  |F(0)| + L^{-[\varphi]j}|DF(0;\1)| + L^{-2[\varphi]j}|D^2F(0;\1,\1)|
  \leq O(1)\|F\|_{T_0(\ell_j)}
  ,
\end{equation}
where $D$ denotes the derivative with respect to $\varphi$ in the direction $\1$ as in \eqref{e:chiSig},
as well as that if
$R^{\varnothing}= R^a|\varphi|^6+  R^g|\varphi|^4+ R^\nu |\varphi|^2 +  R^z (\varphi \cdot (-\Delta \varphi)) +R^u$
then
\begin{equation}
  |R^a_j|, \, |R^z_j|, \,
  L^j |R^g_{j}|  , \,
  L^{2j} |R^{\nu}_{j}|  , \,
  L^{3 j} |R^{ u}_{j}| \leq O(1) \|R^{\varnothing}(B)\|_{T_0(\ell_j)}.
\end{equation}

\begin{theorem}
\label{thm:VK-bulk}
Fix $L$ sufficiently large and $\delta > 0$ sufficiently small.
There are continuous functions $g_0^c,\nu_0^c,z_0^c$ of  $(m^2,a_0) \in [0,\delta)^2$
such that if $(g_0,\nu_0,z_0) = (g_0^c(m^2,a_0),\nu_0^c(m^2,a_0),z_0^c(m^2,a_0))$ then,
for all $j<N$,
\begin{equation}
  \lbeq{ccjbds}
  L^j|g_j|,\, L^{2j}|\nu_j|, L^{3j}|\delta u_j| = O(\vartheta_j a_j),
\end{equation}
and the following remainder bounds hold,
for $B \in \Bcal_j$, $X \in \Ccal_j$, and some (small) $\alpha>0$,
\begin{align}
  \label{e:VKN-Rj-bis}
  \|R^\varnothing_j(B)\|_{T_0(\ell_j)}
  & = O(\vartheta_j^3 a_j^3),
  \\
  \lbeq{Wjbd}
  \|W_j^\varnothing(B)\|_{T_0(\ell_j)} &= O(\vartheta_j^2 a_j^2),
  \\
  \lbeq{Kjbulk}
  \|K_j^\varnothing(X)\|_{T_0(\ell_j)} &= O(\vartheta_j a_j)^{3+\alpha(|\mathcal{B}_j(X)|-2^d)_+}.
\end{align}
The functions $g_0^c,\nu_0^c,z_0^c$ are $O(\csix_0)$, are equal to zero
at $(m^2,0)$, and obey $\ddp{g_0^c}{a_0}, \ddp{\nu_0^c}{a_0},
  \ddp{z_0^c}{a_0} = O(1)$
uniformly in $(m^2,a_0)\in [0,\delta)^2$.
Moreover, all the above bounds hold also for $j=N$ provided $m^2\geq L^{-2N}$.
Finally, the coupling constants
$a_j,g_j,\nu_j,z_j$ and their remainder terms $R_j^\varnothing$
are independent of the volume parameter $N$ provided $N \ge j$.
\end{theorem}

Upper bounds in Theorem~\ref{thm:VK-bulk} are expressed in terms of $a_j$,
so it is important to understand the behaviour of this sequence.  As we discuss
in detail in Section~\ref{sec:barflow} below, $a_j$ obeys the recursion
\begin{equation}
  a_{j+1} = a_j - \beta_j a_j^2 + O(\vartheta_j^3a_j^3)
\end{equation}
with explicit coefficients $\beta_j$.   When $m^2>0$, $a_j$ converges to a constant
which vanishes
logarithmically as $m^2\downarrow 0$, and $a_j\to 0$ when $m^2=0$.
This vanishing of the massless limit goes by the name of \emph{infrared asymptotic freedom}
and is a manifestation of the fact that the RG flows to a Gaussian fixed point.

The next theorem supplements Theorem~\ref{thm:VK-bulk} to permit nonzero values of the
observable fields.  It requires the following definition.
Given $\pp,\qq\in \Lambda$, we define
the \emph{coalescence scale} $j_{\pp \qq}$ to be the unique integer such that
\begin{equation}
\label{e:jabbds}
    \tfrac 12 L^{j_{\pp\qq}} \le |\pp-\qq| < \tfrac 12 L^{j_{\pp\qq}+1},
\end{equation}
namely
$j_{\pp \qq}
    =
    \lfloor
   \log_{L} (2 |\pp - \qq|)
   \rfloor$.
It follows from the finite-range property of $C_j$ that $C_{j;\pp,\qq}= 0$ if $j\le j_{\pp\qq}$.

\begin{theorem}
\label{thm:VK-obs}
Fix $L$ sufficiently large and $\delta > 0$ sufficiently small,
and let $(m^2,a_0) \in [0,\delta)^2$.
Let $(g_0,\nu_0,z_0) = (g_0^c(m^2,a_0),\nu_0^c(m^2,a_0),z_0^c(m^2,a_0))$.
Then, for all $j<N$ and for $x=\pp,\qq$,
the following remainder bounds hold,
for $B \in \Bcal_j$, $X \in \Ccal_j$, and some (small) $\alpha>0$:
\begin{align}
  \lbeq{vlam}
  |R^{\lambda_{x}}_j| & = \1_{j< j_{\pp \qq}} O(\chicCov_{j}  a_{j}^{2} )  ,
  \\
  \lbeq{vq}
  |R^{ q_{x}}_j| &= \1_{j \ge  j_{\pp \qq}}|\pp-\qq|^{-2[\varphi]} 4^{-(j-j_{\pp\qq})} O(\chicCov_{j} a_{j}),
  \\
  \|W^{x}_{j}(B) 
  \|_{T_0(\ell_j)} &=
  L^{-(j\wedge j_{\pp\qq})[\varphi]}2^{-(j-j_{\pp\qq})_+}
  O(\vartheta_j^2 a_j^2),
  \\
  W^{\pp\qq}_j(B)
  &= 0,
  \\
  \lbeq{Kjbd}
  \|K^{x}_j(X)\|_{T_0(\ell_j)}&=
  L^{-(j\wedge j_{\pp\qq})[\varphi]}2^{-(j-j_{\pp\qq})_+}
  O(\vartheta_j a_j)^{2+\alpha(|\mathcal{B}_j(X)|-2^d)_+},
  \\
  \|K^{\pp\qq}_j(X)\|_{T_0(\ell_j)} &=
  L^{-2(j\wedge j_{\pp\qq})[\varphi]}4^{-(j-j_{\pp\qq})_+}
  O(\vartheta_j a_j)^{1+\alpha(|\mathcal{B}_j(X)|-2^d)_+}.
 \label{e:VKN-Kpq-bis}
\end{align}
Moreover, all the above bounds hold also for $j=N$ provided $m^2\geq L^{-2N}$.
Finally, the coupling constants
$\lambda_{x,j},q_{x,j}$ and their remainder terms $R^{\lambda_{x}}_j, R^{q_{x}}_j$
are independent of the volume parameter $N$ provided $N \ge j$.
\end{theorem}

The bounds \eqref{e:VKN-Rj-bis}--\eqref{e:Kjbulk}
and \eqref{e:vlam}--\eqref{e:VKN-Kpq-bis} imply that
  the leading order contributions to the two-point function and susceptibility
  are given by perturbation theory, i.e., by the map $\PT$.
Our main focus is therefore on the analysis of the map $\PT$.
The proof of Theorems~\ref{thm:VK-bulk}--\ref{thm:VK-obs} is discussed
in Appendix~\ref{sec:step-flow}.
It relies significantly on external results adapted from the $4$-dimensional setting.

\subsection{Identification of tricritical point}
\label{sec:tript}

In this section, we identify the tricritical point.

Given $(m^2,a_0)$,
Theorem~\ref{thm:VK-bulk} provides
an initial condition for a
global flow with final conditions $g_\infty=0$ and $\nu_\infty=0$
and with good bounds on
$(V_j,K_j)$.  This allows for an exact computation of $\hat \chi$
in the following corollary to Theorem~\ref{thm:VK-bulk}.

\begin{cor}
\label{cor:suscept}
For  $(m^2,a_0 ) \in (0,\delta)^2$,
\begin{align}
\lbeq{chi-m}
  \hat\chi(m^2,\csix_0,g_0^c(\csix_0,m^2),\nu_0^c(\csix_0,m^2),z_0^c(\csix_0,m^2))
  & = \frac{1}{m^2} .
\end{align}
\end{cor}

\begin{proof}
In this proof, we take $\sigma_\pp=\sigma_\qq=0$.
According to  \refeq{chiZN},
\begin{equation}
\lbeq{chiZN-2}
    \hat\chi_N(m^2,\csix_0,g_0,\nu_0,z_0) =
    \frac{1}{m^2}
    +
    \frac{1}{m^4} \frac{1}{|\Lambda_N|} \frac{D^2 Z_N(0;\1,\1)}{Z_N(0)},
\end{equation}
with $Z_N$ given by the sum in \refeq{ZN2terms}.
It follows from Theorem~\ref{thm:VK-bulk}
that the contributions from $W_N$ and $K_N$ vanish in the limit $N \to \infty$,
and direct computation gives $D^2 e^{-V_N}(0;\1\,\1) = -\nu_N |\Lambda_N|$.
Then \refeq{chi-m} follows from the fact that $\lim_{N\to\infty}\nu_N =0$ by \refeq{ccjbds}.
\end{proof}

On the eight variables $a,g,\nu,m^2,a_0,g_0,\nu_0,z_0$,
we impose the three constraints
\begin{equation}
\label{e:g0gnu0nu-bis}
  \csix_0 = \csix (1+z_0)^3, \quad g_0 = g(1+z_0)^2, \quad \nu_0 = (1+z_0)\nu-m^2
  ,
\end{equation}
of \refeq{g0gnu0nu}, and the three constraints
\begin{equation} \label{e:nu0cz0c}
  g_0 = g_0^c(m^2, \csix_0),
  \quad
  \nu_0 = \nu_0^c(m^2, \csix_0),
  \quad
   z_0 = z_0^c(m^2, \csix_0),
\end{equation}
with $g_0^c,\nu_0^c,z_0^c$ the functions of Theorem~\ref{thm:VK-bulk}.
The next proposition
shows that if we fix $(m^2,a)$ then the other six variables are determined by the constraints.

\begin{prop} \label{prop:changevariables-i}
  There exists $\delta_1>0$
  and continuous functions $(g^*, \nu^*,a_0^*,g_0^*,\nu_0^*,z_0^*)$ of
  $(m^2,\csix)\in [0,\delta_1)^2$,
  such that
  \eqref{e:g0gnu0nu-bis}--\eqref{e:nu0cz0c} hold and
  \begin{gather}
    \label{e:gznustarbd}
    \csix_0^* = \csix + O(\csix^2),
    \quad
    g_0^*,\nu_0^*, z_0^*  = O(\csix).
  \end{gather}
\end{prop}

\begin{proof}
For $(m^2_0,a_0) \in [0,\delta)^2$, set
\begin{equation}
  \lbeq{smg}
  t(m^2, \csix_0) = \frac{\csix_0}{(1+ z_0^c (m^{2}, \csix_{0}))^3}
  .
\end{equation}
By Theorem~\ref{thm:VK-bulk}, $g_0^c,\nu_0^c,z_0^c$ are continuous in
$(m^2,\csix_0) \in [0,\delta)^2$, are $O(a_0)$, and have bounded $a_0$-derivatives.
Thus, with derivatives evaluated at $(m^2,\csix_0)$,
\begin{align} \label{e:sdiff}
  \ddp{t}{\csix_0}
  = \frac{(1+z_0^c)^3 - 3\csix_0 (1+z_0^c)^2 \ddp{z_0^c}{\csix_0}}{(1+z_0^c)^6}
  = 1+O(\csix_0) > 0.
\end{align}
For sufficiently small $\delta>0$, $t$ is therefore a strictly increasing
continuous function of $\csix_{0} \in [0,\delta)$ such that
$|t (m^2,u) - t (m^2,v)| \ge (1 - O (\delta)) |u - v|$
and hence, for $m^2,\ginfty$ fixed,
$t(m^2,\cdot)$ is a continuously invertible map from $[0,\delta)$ onto
the interval $[0,t(m^2,\delta))$.  We denote the inverse map as $\csix_0^*$,
set
\begin{align}
  g_0^*(m^2,\csix) &= g_0^c(m^2,\csix_0^*(m^2,\csix))
  ,
  \quad
  \nu_0^*(m^2,\csix) = \nu_0^c(m^2,\csix_0^*(m^2,\csix))
  ,
  \quad
  z_0^*(m^2,\csix) = z_0^c(m^2,\csix_0^*(m^2,\csix))
  ,
\end{align}
and define
\begin{equation}
  \label{e:mu-m-def}
  g^*(m^2 , \csix) = \frac{g_0^*}{(1+z_0^*)^2},
  \qquad
  \nu^*(m^2 , \csix) = \frac{\nu_0^* + m^2}{1+z_0^*}.
\end{equation}
Since $\nu_0^c,z_0^c,\csix^*$ are continuous,
it is also the case that $g_0^*,\nu_0^*,z_0^*,\nu^*$
are continuous.  It is immediate
that \eqref{e:g0gnu0nu-bis}--\eqref{e:nu0cz0c} hold,
and also that \eqref{e:gznustarbd} holds.

Let $\delta_1 = \frac 12 \delta$.  By \refeq{smg},
$[0,\delta_1)$ lies in the
intersection over $m^{2}>0$ of the intervals $[0,t(m^2,\delta))$.
Therefore, $a=t(m^2,a_0)$ can be solved for $a_{0}$ as a
function $a_0^*(m^{2},a)$ for $a \in [0,\delta_1)$ and $a_{0}^*$ is
continuous in $a$ for $m^2$ fixed.
To see that $a_{0}^*$ is jointly continuous in $(m^2,a)$, it suffices
to show that if $(\hat m^2,\hat a) \rightarrow (m^2,a)$ then $\hat
a_{0} \rightarrow a_{0}$, where $\hat a_{0},a_{0}$ solve $t(\hat
m^2,\hat a_{0}) - \hat a = 0 = t(m^2,a_{0}) - a$.  This follows from
$(1 - O(a_{*})) |\hat a_{0} - a_{0}| \le |t (\hat m^2,\hat a_{0}) - t
(\hat m^2,a_{0})| = |(t (m^2,a_{0}) - t (\hat m^2,a_{0})) + (\hat a-
a)| \rightarrow 0$, since $t(\cdot,a_0)$ is continuous by \refeq{smg}
and the continuity of $z_0^c$.
\end{proof}

With the continuous functions produced in Proposition~\ref{prop:changevariables-i},
we see from  \refeq{chi-m} that
\begin{equation}
  \chi(\csix,g^* ,\nu^* )
  = (1+z_0^*) \hat\chi(m^2,a_0^*, g_0^*,\nu_0^*,z_0^*)
    = (1+z_0^* )\frac{1}{m^2}.
\end{equation}
In particular, the susceptibility $\chi(a,g^*, \nu^* )$ is finite if $m^2 > 0$,
whereas
$\chi(\csix,g^*,\nu^*) \to \infty$ as $m^2 \downarrow 0$.
The divergence of the susceptibility, together with the fact that the RG fixed
point is Gaussian, leads us to define the \emph{tricritical point} as
\begin{equation} \label{e:nucinf}
    (g_c(\csix),\nu_c(\csix)) = (g^*(0,\csix),\nu^*(0,\csix)).
\end{equation}
We will see in Section~\ref{sec:pf-main-n} that
$(g^*(m^2,a),\nu^*(m^2,a))$ defines the curve of Theorem~\ref{thm:main-n},
parametrised by $m^2$. The geometry of the curve is of interest,
but we do not investigate it in this paper.

\section{The map $\PT$ and the approximate flow}
\label{sec:PT}

In this section, we define the perturbative map $\PT$ and its simplification
called the \emph{approximate flow}.
In particular, we incorporate improvements to the treatment of
the map $\PT$ used in \cite{BBS-rg-pt,BBS-phi4-log}, and extend it to include
a $|\varphi|^6$ term.  The improvements include a more systematic treatment
of the change of variables
(transformation) used in \cite{BBS-rg-pt}, as well as the use of general estimates
for coefficients arising in flow equations
rather than detailed individual estimates based on explicit formulas.
The main result of this section is Proposition~\ref{prop:barflow}, which provides
the approximate flow.

We use the notation appropriate for $n \ge 1$.
The relevant ($1,\tau,\tau^2$) and marginal
($\tau^3,\tau_{\Delta},\tau_{\nabla\nabla}$)
bulk monomials obeying Euclidean and $O(n)$ symmetry are:
\begin{gather}
\lbeq{monomials}
  1, \quad \tau = \half |\varphi|^2, \quad \tau^2 = \tfrac14 |\varphi|^4,
  \quad \tau^3 = \tfrac18 |\varphi|^6
  ,
  \\
  \tau_{\Delta} = \half \varphi \cdot (-\Delta \varphi)
  , \quad
  \tau_{\nabla\nabla} =
  \tfrac14 \sum_{e\in\Z^d:|e|_1=1} \nabla^e\varphi \cdot \nabla^e \varphi
  .
\end{gather}
The following complex vector spaces of polynomials play a role:
\begin{align}
    \Vcal^\varnothing
    & = \{ a\tau^3 + g \tau^2 + \nu \tau  + z\tau_\Delta : a,g,\nu,z  \in \C \},
    \\
    \Ucal^\varnothing
    & = \{ V + y \tau_{\nabla\nabla} + u : V \in \Vcal^\varnothing, \, y,u \in \C \},
    \\
\lbeq{Vcaldef}
    \Vcal
    & = \{ V -\lambda_{\pp}\sigma_\pp (\onehat\cdot \varphi)\1_{\pp}
     -\lambda_{\qq}\sigma_\qq (\onehat\cdot \varphi)\1_{\qq}
       :
     V \in \Vcal^\varnothing, \,\lambda_{\pp},\lambda_{\qq}  \in \C \},
    \\
    \Ucal
    & = \{ U
    - \lambda_{\pp}\sigma_\pp (\onehat\cdot \varphi)\1_{\pp}
    - \lambda_{\qq}\sigma_\qq (\onehat\cdot \varphi)\1_{\qq}
    - \tfrac 12( q_{\pp}\1_\pp +  q_{\qq}\1_\qq )\sigma_\pp  \sigma_\qq
    : U \in \Ucal^\varnothing, \, \lambda_{\pp},\lambda_{\qq} ,q_{\pp}\, q_{\qq}
    \in \C \}.
\end{align}
The field $\varphi$ is evaluated at a point $x\in \Lambda$, and here $\1_\pp$
represents the Kronecker delta $\1_\pp(x)= \1_{\pp=x}= \delta_{\pp,x}$.
Given $X \subset \Lambda$, we also define, e.g.,
\begin{equation}
\label{e:Vcalesig}
    \Vcal(X) = \{V(X) = \textstyle{\sum_{x\in X}} V_x : V \in \Vcal \}.
\end{equation}
The counterparts of these monomials and spaces for $n=0$ are
modified as in \cite{BBS-rg-pt}.

\subsection{Localisation}

Given $X \subset \Lambda$,
the localisation operator is a linear map
$\LT_X$ which projects $\Ncal$ onto a subspace of
polynomials consisting of relevant and marginal monomials summed over $X$,
essentially as a Taylor expansion.
Since $\LT_X$ preserves $O(n)$ symmetry, in practice it maps into $\Ucal(X)$.
The definition and properties of $\LT$ are developed in detail in \cite{BS-rg-loc};
it involves parameters which we specify in Section~\ref{sec:loc}.
The following example
gives 
the action of $\LT_x$ with these parameters when $X=\{x\}$ is a single point;
this is all that is required for the rest of Section~\ref{sec:PT}.

\begin{example}
\label{ex:Loc}
For notational simplicity, suppose that the number of field components is $n=1$.

\smallskip\noindent (i)
If $r+s$ is even then
\begin{equation}
    \LT_x \varphi_x^r\varphi_y^s =
    \begin{cases}
    0 &  (r+s>6) \\
    \varphi_x^{r+s} &  (r+s=4,6).
    \end{cases}
\end{equation}
Also, $\LT_x \varphi_x^r = \varphi_x^r$ for $r \le 6$, whereas $\LT_x \varphi_x^r = 0$
for $r > 6$.

\smallskip\noindent (ii)
Suppose that $p: \Lambda \to \R$ satisfies $p_x=0$ if $|x|> \half \diam{\Lambda} =\half L^N$
and that, for some $p^{(**)} \in \R$,
\begin{equation}
  \label{e:qprop1}
  \sum_{x \in \Lambda} p_x x_{i} = 0,
  \quad\quad
  \sum_{x \in \Lambda} p_x x_{i}x_{j}
  = p^{(**)} \delta_{i,j},
  \quad\quad\quad
  i,j \in \{1,2,3 \}.
\end{equation}
Then, as in \cite[Section~1.5]{BS-rg-loc} or \cite[(5.29)--(5.30)]{BBS-rg-pt},
with $\tau_{xy}= \frac 12 \varphi_x\varphi_y$ and $p^{(1)}=\sum_x p_x$,
\begin{align}
\label{e:LTF3}
  \LT_{x}
  \left(
    \sum_{y \in \Lambda} p_{x-y} \tau_{y}
  \right)
  &=
  p^{(1)}\tau_{x}
  +
  p^{(**)} (\tau_{\nabla\nabla,x}-\tau_{\Delta,x}),
  \\
\label{e:LTF4}
  \LT_{x}
  \left(
    \sum_{y \in \Lambda} p_{x-y} \tau_{xy}
  \right)
  &=
  p^{(1)}\tau_x
  +
  p^{(**)} \tau_{\Delta,x}.
\end{align}
\end{example}

\subsection{Definition of the map \texorpdfstring{$\PT$}{PT}}
\label{sec:PTdef}

In this section, we define the quadratic map $\PT_j: \Ucal \to \Ucal$
(``$\PT$'' stands for ``perturbation theory'').
It is designed in such a way
that if $Z_j$ is represented perturbatively as $Z_j \approx
e^{-U_j(\Lambda)}$ for a polynomial
$U_j \in \Ucal$, then the map $Z_j \mapsto Z_{j+1}$ can be approximated by
the map $U_j \mapsto \PT_j(U_j)$.  This is discussed in detail in \cite[Section~2]{BBS-rg-pt}.
We use the notation here for
$n \ge 1$; the adaptation to $n=0$ can be found in \cite{BBS-rg-pt}.

Given a $\Lambda \times\Lambda$ matrix $C$, we define a linear operator on
sufficiently differentiable complex-valued functions of $\varphi$ by
\begin{equation}
\label{e:LapC}
    \Lcal_C =
    \frac 12
    \sum_{i=1}^n
    \sum_{u,v \in \Lambda}
    C_{u,v}
    \frac{\partial}{\partial \varphi_{u}^i}
    \frac{\partial}{\partial \varphi_{v}^i}.
\end{equation}
For a polynomial $A$ in the field, $\Ex_C\theta A = e^{\Lcal_C}A$,
where the exponential is defined by power series expansion which terminates when applied
to a polynomial (see \cite[Lemma~4.2]{BS-rg-norm}).
For polynomials $A,B$ in the field, we define
\begin{equation}
  \label{e:FCAB}
    F_{C}(A,B)
    = e^{\Lcal_C}
    \big(e^{-\Lcal_C}A\big)
    \big(e^{-\Lcal_C}B\big) - AB
    .
\end{equation}
As in \cite[Lemma~5.6]{BBS-rg-pt}, $F$ can be evaluated using
\begin{equation}
\label{e:Fexpand1Vpt}
    F_{C}(A_x,B_y) = \sum_{k=1}^{\deg A\wedge \deg B}  \frac{1}{k!}
    \sum_{i_1,\ldots,i_k=1}^n
    \sum_{u_l,v_l \in \Lambda}
    \left(\prod_{l=1}^k C_{u_l,v_l}\right)
    \frac{\partial^k A_x}{\partial \varphi_{u_1}^{i_1}\cdots \partial \varphi_{u_k}^{i_k}}
    \frac{\partial^k B_y}{\partial \varphi_{v_1}^{i_1}\cdots \partial \varphi_{v_k}^{i_k}}.
\end{equation}

Let $w_0=0$.  For $j \ge 1$, and for $C_i$ the terms in the covariance decomposition
\refeq{ZdCj}, let
\begin{equation}
\lbeq{wdef}
    w_j = \sum_{i=1}^j C_i.
\end{equation}
The range of $w_j$ is the same as that of $C_j$, namely $\frac 12 L^j$.
For $U \in \Ucal$ and $X \subset \Lambda$, we set
\begin{equation}
  \label{e:WLTF}
  W_j(U,X) = \frac 12 \sum_{x\in X} (1-\LT_{x}) F_{w_j}(U_x,U(\Lambda)).
\end{equation}

The map $\PT_j:\Ucal \to \Ucal$ is then defined by
\begin{equation}
  \lbeq{Vptdef}
  \PT_j(U) = e^{\Lcal_{C_{j+1}}} U - P_j(U)
\end{equation}
with
\begin{align}
\label{e:PdefF}
    P_{x,j}(U)
    &=
    \LT_x \left(
    e^{\Lcal_{C_{j+1}}} W_j(U,x)
    + \frac 12
    F_{C_{j+1}}
    (e^{\Lcal_{C_{j+1}}} U_x,e^{\Lcal_{C_{j+1}}} U(\Lambda))
    \right).
\end{align}
By translation invariance, $P_{x,j}(U)$
defines a local polynomial with coefficients independent of $x$.
According to \cite[Lemma~5.5]{BBS-rg-pt}, an equivalent alternate formula for
$P_{x,j}(U)$
is
\begin{align}
\label{e:Palt0}
    P_{x,j}(U)
    &=
    \frac 12
    \left(
    \LT_{x}
    F_{w_{j+1}} (e^{\Lcal_{C_{j+1}}} U_x,e^{\Lcal_{C_{j+1}}} U(\Lambda)) -
    e^{\Lcal_{C_{j+1}}} \LT_{x}  F_{w_j} (U_x,U(\Lambda))
    \right)
    .
\end{align}

\subsection{Linear term}
\label{sec:linearterm}

Throughout Sections~\ref{sec:linearterm}--\ref{sec:changeofvariables}, we study only on the bulk,
and return to
observables in Section~\ref{sec:flowobs}.

According to \refeq{Vptdef}, the linear term in the map $\PT$ is given by
$U\mapsto e^{\Lcal_{C_{j+1}}}U$.  In the following lemma, we compute
this linear map, for $U \in \Ucal^\varnothing$ and for an arbitrary covariance $C$.
The matrix is with respect to the representation of $U = a\tau^3+g\tau^2+\nu\tau
+ y \tau_{\nabla\nabla} + z \tau_\Delta + u \in \Ucal^\varnothing$ as
\begin{equation}
    U = (u,\nu,g,a,y,z).
\end{equation}

\begin{lemma}
\label{lem:eLcal}
Let $n \ge  0$.
The linear map $e^{\Lcal_C}$ on $\Ucal$ has matrix representation,
with $c=C_{0,0}$ and $c_\Delta = \Delta C_{0,0}$,
\begin{align}
\lbeq{Wickconstants}
e^{\Lcal_C} &=  \left(\begin{smallmatrix}
1& \frac 12 nc & \frac{1}{4} n(n+2)c^2 & \frac{3}{32} n(n+2)(n+4)c^3
& -\frac 12 nc_\Delta & -\frac 12 nc_\Delta
 \\
0  & 1 & (n+2)c & \frac{3}{4}(n+4)(n+2) c^2 & 0 & 0 \\
0 & 0  & 1 & \frac{3}{2}(n+4)c & 0 & 0 \\
0 & 0 & 0 & 1& 0 & 0\\
0 & 0  & 0 & 0& 1 & 0 \\
0 & 0 & 0 & 0 & 0 & 1
\end{smallmatrix}\right).
\end{align}
\end{lemma}

\begin{proof}
The exponential of $\Lcal=\Lcal_{C}$ is defined by expansion in Taylor series,
which gives, for $U \in \Ucal$,
\begin{equation}
    e^{\Lcal}  U =
    \left(1 + \Lcal + \frac {1}{2!} \Lcal^2 + \frac {1}{3!} \Lcal^3 \right) U.
\end{equation}
Suppose first that $n \ge 1$. Differentiation gives
\begin{gather}
    \Lcal |\varphi|^2  = nc,
    \qquad
    \Lcal |\varphi|^4  = \frac 12 4(n+2)c|\varphi|^2,
    \qquad
    \Lcal |\varphi|^6  = \frac 12 6(n+4)c|\varphi|^4,
    \\
    \Lcal (\varphi \cdot \Delta \varphi)  =
    \frac 12 \sum_{|e|_1=1} \Lcal (\nabla^e \varphi \cdot \nabla^e \varphi)
    =-   n \Delta C_{0,0} ,
\end{gather}
from which the desired formula can be obtained after some algebra.
Similar computations in the supersymmetric setting give the result for $n=0$,
as in \cite{BBS-rg-pt}.
\end{proof}

\subsection{Dimensionless form of the perturbative flow}

We define $D=\{0,1,2,3,\nabla\nabla,\Delta\}$ as a set whose elements
are representatives of the monomials
$1,\tau,\tau^2, \tau^3, \tau_{\nabla\nabla},\tau_\Delta$.
We write these monomials as $M^i$ for $i \in D$.
The \emph{dimension} of $i\in\{0,1,2,3\}$ is defined to be $[i]=i$, and $[\nabla\nabla]=[\Delta]=3$.

We rewrite the coupling constants as
\begin{equation}
  \gamma_0 =u, \quad \gamma_1 =\nu, \quad \gamma_2 = g, \quad \gamma_3 = \csix,
  \quad \gamma_{\nabla\nabla}=y, \quad \gamma_\Delta=z.
\end{equation}
By \refeq{Vptdef}, the perturbative flow equations,
which express the coupling constants of $\PT(U)$
in terms of those of $U$, have the form
\begin{equation}
\lbeq{flow-initial}
  \gamma_{i,\pt}
  =  \sum_{p\in D}
  \alpha_i^p \gamma_p
  -
  \sum_{p,q\in D}
  \alpha_i^{pq}\gamma_p\gamma_q.
\end{equation}
The linear coefficients $\alpha_i^p$ are matrix elements of \refeq{Wickconstants}.
The quadratic coefficients $\alpha_i^{pq}$ are our main concern in the rest of this section.
They can be computed exactly as in \cite{BBS-saw4-log,BBS-phi4-log}, but mostly we do
not need exact values here.
The following lemma obtains estimates much more efficiently than those obtained
from exact formulas in \cite{BBS-rg-pt,BBS-phi4-log}.
Recall that $\chicCov_j$ is defined in \refeq{chicCovdef}.

\begin{lemma}
\label{lem:greekbds}
For $i,p,q\in D$, the coefficients $\alpha_i^{p},\alpha_i^{pq}$
in \refeq{flow-initial} obey the estimates
\begin{align}
    \alpha_i^i & = 1,
    \\
    \alpha_0^p  & = O(\chicCov_j L^{-j[p]})     \quad\quad      (p \neq 0) ,
    \\
    \alpha_1^p  & = O(\chicCov_j L^{-j([p]-1)})     \quad      (p =2,3) , \\
    \alpha_2^3  & = O(\chicCov_j L^{-j(3-2)})     ,
\\
  \alpha_i^{pq} &= O_L(\chicCov_j L^{j(3+[i]-[p]-[q])}),
\end{align}
and all $\alpha_i^p$ not listed above are equal to zero.
\end{lemma}

\begin{proof}
For the linear terms, we can read off the coefficients from Lemma~\ref{lem:eLcal},
and the bound follows from the estimate \refeq{scaling-estimate-theta} on the covariance.

For the quadratic terms, we use the $T_0(\ell)$-seminorm  (defined
in \cite[Section~2.3]{BBS-phi4-log} for $n \ge 1$ and in \cite[Section~6.3]{BBS-saw4-log}
for $n=0$) with parameter
\begin{equation}
\lbeq{elldef}
    \ell_j = \ell_0 L^{-j[\varphi]}
    = \ell_0 L^{-j/2},
\end{equation}
with $\ell_0$ a (large) $L$-dependent constant.
If $B$ is a $j$-block, then a calculation gives
\begin{equation}
\lbeq{mondim}
    \|M^r(B)\|_{T_0(\ell)} \asymp_L L^{j(3-[r])},
\end{equation}
where the notation $\asymp_L$ indicates
upper and lower bounds with constants that may depend on $L$.
We apply \cite[Proposition~4.10]{BS-rg-IE} and \refeq{mondim} to see that
\begin{equation}
  \|P(M^p(B),M^q(B))\|_{T_0(\ell)} = O_L(\chicCov) L^{j(3-[p])}L^{j(3-[q])}.
\end{equation}
Since, by definition,
\begin{equation}
  P(M^p(B),M^q(B))
  =
  \sum_{i\in D} \alpha^{pq}_{i}M^i(B)
  ,
\end{equation}
it follows from the fact that $\|\alpha^{pq}_{i}M^i (B)\|_{T_0(\ell)} \le
\|\sum_{i\in D} \alpha^{pq}_{i}M^i(B)\|_{T_0(\ell)}$ (as in \cite[(3.4)]{BS-rg-IE}) that
\begin{equation}
  \|\alpha^{pq}_{i}M^i (B)\|_{T_0(\ell)} \le O_L(\chicCov) L^{j(6-([p]+[q]))},
\end{equation}
so that
\begin{equation}
  |\alpha^{pq}_{i}| \le O_L(\chicCov) L^{j(6-([p]+[q]))} L^{-j(3-[i])}
  = O_L(\chicCov)  L^{j(3+[i]-[p]-[q])} .
\end{equation}
This completes the proof.
\end{proof}

We now rescale to dimensionless variables, as follows.
Let
\begin{equation}
  \hat\gamma_i = L^{(3-[i])j}\gamma_i , \qquad
  \gamma_i = L^{-(3-[i])j}\hat\gamma_i.
\end{equation}
We define dimensionless coefficients,
with all $\hat\alpha_i^{p},\hat\alpha_i^{pq}$
bounded by $\chicCov$ (except $\hat\alpha_i^i=1$),  by
\begin{equation}
\lbeq{betahatbds}
    \hat\alpha_i^p = L^{j([p]-[i])}\alpha_i^p,
    \qquad
    \hat\alpha_i^{pq} = L^{j(-3-[i]+[p]+[q])})\alpha_i^{pq} .
\end{equation}
Then we can rewrite the original perturbative flow equations \refeq{flow-initial}
in dimensionless form as
\begin{equation}
\lbeq{hatgamflow}
  \hat\gamma_{i,\pt}
  = L^{3-[i]}
  \left(
  \sum_{p\in D}
  \hat\alpha_i^p
  \hat\gamma_p
  -
  \sum_{p,q\in D}
  \hat\alpha_i^{pq}\hat\gamma_p\hat\gamma_q
  \right)
  \qquad (i\in D)
  .
\end{equation}

\subsection{Change of variables}
\label{sec:changeofvariables}

In this section, we make a change of variables and transform the dimensionless
perturbative flow equations \refeq{hatgamflow} into a triangular form.
This is achieved in Proposition~\ref{prop:barflow}.  A related transformation was
used in \cite[Section~4.2]{BBS-rg-pt} in a more ad hoc manner.  Here we present
the transformation in a systematic way.

Let $U \in \Ucal^\varnothing$.
We write $F=F_w$, $F_+=F_{w+C}$, and $\Lcal_+ = \Lcal_{w+C}$.
By \refeq{Palt0}, the quadratic term in $\PT(U)$ is
\begin{equation}
    P_x(U) =
    \frac 12
    \left(
    \LT_x F_{+}(e^{\Lcal_{+}}U_x; e^{\Lcal_{+}}U(\Lambda))
    -
    e^{\Lcal_{+}}
    \LT_x F(U_x; U(\Lambda))
    \right).
\end{equation}
Let $Q(U,U)= \frac 12 \LT_x F(U_x; U(\Lambda))$.  There are coefficients $\kappa_i^{pq}$
($i,p,q\in D$) such that
\begin{equation}
\lbeq{kapdef}
  Q (M_x^{p},M^{q}(\Lambda))
  = \sum_{i\in D} \kappa_i^{pq} M_x^{i}.
\end{equation}
Rescaled versions of the coefficients are defined, as in \refeq{betahatbds}, by
\begin{equation}
\lbeq{kaphatdef}
  \hat\kappa_i^{pq} = L^{j(-3-[i]+[p]+[q])}\kappa_i^{pq} .
\end{equation}
The coefficients $\hat\kappa_i^{pq}$ may be bounded or unbounded in the scale $j$, and we define
\begin{equation}
    S^{\rm bd} = \{(i,p,q) : \hat\kappa_i^{pq}= O(1)\},
\end{equation}
and, for $U = \sum_{i\in D} \gamma_i M^i$, set
\begin{equation}
    Q^{\rm bd}(U,U) = \sum_{(i,p,q)\in S^{\rm bd}} \kappa_i^{pq} \gamma_p \gamma_q M_x^i,
    \qquad
    Q^{\rm div}(U,U) = Q(U,U) - Q^{\rm bd}(U,U).
\end{equation}
Thus we have divided $Q$
into its bounded and unbounded terms as $Q=Q^{\rm bd}+Q^{\rm div}$, so
\begin{align}
    P_x(U)
    &=
    Q_{+}(e^{\Lcal_{+}}U; e^{\Lcal_{+}}U)
    -
    e^{\Lcal_{+}}Q(U,U)
    \nnb & =
    [Q_{+}^{\rm div}(e^{\Lcal_{+}}U; e^{\Lcal_{+}}U)
    -
    e^{\Lcal_{+}}Q^{\rm div}(U,U) ]
    \nnb & \quad +
    [Q_{+}^{\rm bd}(e^{\Lcal_{+}}U; e^{\Lcal_{+}}U)
    -
    e^{\Lcal_{+}}Q^{\rm bd}(U,U) ]
    .
\end{align}
We write the map $\PT$ as $\varphi_{\pt}$, set $\Upt = \varphi_{\pt}(U)$,
and rewrite the equation $\Upt = e^{\Lcal_+}U - P$ as
\begin{align}
\lbeq{transform1}
    \varphi_{\pt}(U) + Q^{\rm bd}_+(e^{\Lcal_{+}}U; e^{\Lcal_{+}}U)
    =
    e^{\Lcal_+}(U+Q^{\rm bd}(U,U))
    - [Q_{+}^{\rm div}(e^{\Lcal_{+}}U; e^{\Lcal_{+}}U)
    -
    e^{\Lcal_{+}}Q^{\rm div}(U,U) ].
\end{align}

Next, we define a transformation $T:\Ucal\to\Ucal$ by
\begin{equation}
    T(U) = U+Q^{\rm bd}(U,U).
\end{equation}
Note that $T$ is equal to the identity map plus a quadratic part.
It follows from \refeq{transform1} that
\begin{align}
    &\varphi_{\pt}(U) + Q^{\rm bd}_+(\varphi_{\pt}(U); \varphi_{\pt}(U))
    \\ \nonumber & \quad =
    e^{\Lcal_+}(U+Q^{\rm bd}(U,U))
    - [Q_{+}^{\rm div}(e^{\Lcal_{+}}T(U)U; e^{\Lcal_{+}}T(U))
    -
    e^{\Lcal_{+}}Q^{\rm div}(T(U),T(U)) ]
    +O(U^3),
\end{align}
and hence
\begin{align}
    (T_+\circ \varphi_{\pt}\circ T^{-1})(U)
    =
    e^{\Lcal_+}U
    - [Q_{+}^{\rm div}(e^{\Lcal_{+}}U; e^{\Lcal_{+}}U)
    -
    e^{\Lcal_{+}}Q^{\rm div}(U,U) ]
    +O(U^3).
\end{align}
The \emph{approximate flow} is defined by dropping the error term in the above, which yields
\begin{equation}
    \bar U_+ =
    e^{\Lcal_+}\bar U
    - [Q_{+}^{\rm div}(e^{\Lcal_{+}}\bar U; e^{\Lcal_{+}}\bar U)
    -
    e^{\Lcal_{+}}Q^{\rm div}(\bar U,\bar U) ].
\end{equation}

The following lemma identifies several coefficients whose indices belong to $S^{\rm bd}$.
The proof of the lemma shows that many of the options in \refeq{kipD} are in fact
zero, but since we do not need to know they are zero, we state the weaker bounds for simplicity.
It is not necessary to transform the variables $u,y$ so we omit them from the
following discussion.

\begin{lemma}
\label{lem:triangular}
For $i,p,q\in \{1,2,3,\Delta\}$, the coefficients obey
\begin{align}
\lbeq{kipq}
    \hat\kappa_i^{pq} & =O(1) \qquad (p+q<3+i , \; i,p,q\in \{1,2,3\}),
    \\
\lbeq{k113}
    \hat\kappa_1^{13} = \hat\kappa_1^{31} & = O(1),
    \\
\lbeq{kDpq}
    \hat\kappa_\Delta^{pq} & = O(1)
    \qquad (p,q\in \{1,2,3\} \; \text{except} \; (p,q)=(3,3)),
    \\
\lbeq{kipD}
    \hat\kappa_i^{\Delta p}=\hat\kappa_i^{p\Delta} & = O(1)
    \qquad (i,p\in \{1,2,3,  
    \Delta\}) .
\end{align}
\end{lemma}

Before proving the lemma, we discuss its important consequence that
the approximate flow is triangular.

\begin{prop}
\label{prop:barflow}
The approximate flow has the following form,
with all coefficients $\bar\beta=O(\chicCov)$:
\begin{align}
\lbeq{abarflow}
    \bar\gamma_{3,+} & = \bar\gamma_3 - \bar\beta_3^{33} \bar\gamma_{3}^2,
    \\
\lbeq{zbarflow}
    \bar\gamma_{\Delta,+} & = \bar\gamma_{\Delta}
    - \bar\beta_\Delta^{33}\bar\gamma_{3}^2,
    \\
    \bar\gamma_{2,+} & =
    L\left( \bar\gamma_2 + \bar\beta_2^3\bar\gamma_3
    - \bar\beta_2^{23}\bar\gamma_2\bar\gamma_3 - \bar\beta_2^{33}\bar\gamma_3^2 \right),
    \\
\lbeq{nubarflow}
    \bar\gamma_{1,+} & =
    L^2 \left( \bar\gamma_{1}
    + \bar\beta_1^2\bar\gamma_2  + \bar\beta_1^3\bar\gamma_3
    - \bar\beta_1^{22}\bar\gamma_2^2 -
    \bar\beta_1^{23}\bar\gamma_2\bar\gamma_{3}
    -
    \bar\beta_1^{33}\bar\gamma_{3}^2
    \right).
\end{align}
\end{prop}

\begin{proof}[Proof assuming Lemma~\ref{lem:triangular}]
The linear terms on the right-hand sides of \refeq{abarflow}--\refeq{nubarflow}
have the desired form by Lemma~\ref{lem:greekbds}, so
we only need to study the quadratic terms.
By Lemma~\ref{lem:greekbds}, the quadratic coefficients are all $O(\chicCov)$.

It follows from \refeq{kipq} that there are no $\bar\gamma_1$ or $\bar\gamma_2$ terms in the
$\bar\gamma_3$ equation, that there are no $\bar\gamma_1$ terms in the $\bar\gamma_2$ equation,
and that there
is no $\bar\gamma_1\bar\gamma_2$ or $\bar\gamma_1\bar\gamma_1$ term in the
$\bar\gamma_1$ equation.

It follows from \refeq{k113} that there is no $\bar\gamma_1\bar\gamma_3$ term in the
$\bar\gamma_1$ equation.

It follows from \refeq{kDpq} that the $\bar\gamma_\Delta$
equation can only
have a $\bar\gamma_3^2$ term, among $\bar\gamma_1, \bar\gamma_2,\bar\gamma_3$.

It follows from \refeq{kipD} that no $\bar\gamma_\Delta$
terms occur in any equation.
\end{proof}

Concerning the coefficients $\bar\beta$, by Lemma~\ref{lem:eLcal} the linear ones are
\begin{gather}
\lbeq{beta1-23}
    \bar\beta_{1,j}^2 = (n+2)L^jC_{j+1;0,0},
    \quad
    \bar\beta_{1,j}^3 = \frac 34 (n+4)(n+2)L^{2j}C_{j+1;0,0}^2,
    \\
\lbeq{beta_2-3}
    \bar\beta_{2,j}^3 = \frac 32 (n+4) L^{ j}C_{j+1;0,0}
    .
\end{gather}
For the quadratic coefficients, we extend \refeq{wdef} by defining, for
integers $j \ge 0$ and $k \ge 1$,
\begin{equation}
\lbeq{wndef}
  w_j = \sum_{k = 1}^{j} C_k,
  \qquad
  w^{(k)}_j = \sum_{x\in \Z^d} w_{j;0x}^k,
  \qquad
  w^{(k,**)}_j = \sum_{x \in \Z^d} x_1^2 w_{j;0x}^k.
\end{equation}
Bounds on these quantities are given in Lemma~\ref{lem:w}.
Two marginal coefficients play a specific role, which can
be computed directly (including the case $n=0$) as
    \begin{align}
    \lbeq{b333def}
      \bar\beta_3^{33} &= (18n+132) \delta[w^{(3)}],
      \qquad
      \bar\beta_2^{23} = (12n+48) \delta[w^{(3)}],
    \end{align}
where we write in general $\delta[f] = f_{j+1}-f_j$.
For later use, we define $p_2$ as the ratio
    \begin{equation}
    \lbeq{p2def}
       p_2 = \frac{\bar\beta_2^{23}}{\bar\beta_3^{33}} = \frac{2(n+4)}{3n+22}.
    \end{equation}
By Proposition~\ref{prop:barflow}, $\delta[w^{(3)}]$ is bounded,
but in fact this results from a difference of unbounded
terms.
Although we do not need it, explicit computation also gives
    \begin{align}
      \bar\beta_1^{22} &= (2n+4) \delta[w^{(3)}],
      \qquad
\lbeq{betaDelta33}
      \bar\beta_{\Delta}^{33} = -\frac{9}{4} (n+2)(n+4)
      \delta[
      w^{(5,**)}
      ] + \text{bounded}
      .
    \end{align}

\begin{proof}[Proof of Lemma~\ref{lem:triangular}]
For notational simplicity, suppose that the number of field components is $n=1$.
Recall from \refeq{kaphatdef} that $\hat\kappa_i^{pq} = L^{j(-3-[i]+[p]+[q])}\kappa_i^{pq}$,
where $\kappa_i^{pq}$ is the coefficient appearing in
\begin{equation}
\lbeq{kapdef-bis}
  Q (M_x^{p},M^{q}(\Lambda)) = \sum_y \frac 12 \LT_xF(M_x^p,M_y^q)
  = \sum_{i\in D} \kappa_i^{pq} M_x^{i}.
\end{equation}

\smallskip\noindent \emph{Proof of \refeq{kipq}.}
Suppose first that $p,q \in \{1,2,3\}$ and that $p+q-i<3$.
By the formula for $F$ in \refeq{Fexpand1Vpt}, there are $c_{p,q,k}$ such that
\begin{align}
    \LT_x F(\varphi_x^{2p},\varphi_y^{2q})
    & =
    \sum_{k=1}^{2p \wedge 2q}
    c_{p,q,k} w_{xy}^k \LT_x \varphi_x^{2p-k}\varphi_y^{2q-k}.
\end{align}
By Example~\ref{ex:Loc}, if $p+q-k \ge 2$ then
\begin{equation}
    \LT_x \varphi_x^{2p-k}\varphi_y^{2q-k} = \1_{p+q-k\le 3} \,  \varphi_x^{2(p+q-k)},
\end{equation}
and hence, for $i=p+q-k$ (which entails $k<3$ by hypothesis),
\begin{equation}
\lbeq{kapbd}
    \hat\kappa_i^{pq} \propto L^{j(-3-i+p+q)}w_j^{(k)} = O(1),
\end{equation}
since $w_j^{(k)} \le O(L^{j(3-k)})$ by Lemma~\ref{lem:w}.
If instead $p+q-k=1$ then it follows from Example~\ref{ex:Loc} that
\begin{equation}
\lbeq{LTwnstars}
    \LT_x \sum_y \varphi_x^{2p-k}w_{xy}^k\varphi_y^{2q-k}
    = \varphi_x^2 w^{(k)} + \1_{2q-k>0} \varphi^2_{\Delta,x} w^{(k,**)},
\end{equation}
where $\varphi^2_{\Delta,x}$ represents a linear combination of $\tau_{\Delta,x}$
and $\tau_{\nabla\nabla,x}$.
The first term in \refeq{LTwnstars} yields $\hat \kappa_1^{pq}$,
which is $O(1)$ as in \refeq{kapbd}.

\smallskip\noindent \emph{Proof of \refeq{kDpq}.}
The second term in \refeq{LTwnstars}
gives rise to $\hat\kappa_\Delta^{pq}$
and $\hat\kappa_{\nabla\nabla}^{pq}$,
and the relation $p+q=k+1$ implies
that $k<5$ if we exclude $(p,q)=(3,3)$.
Therefore, by Example~\ref{ex:Loc} and by Lemma~\ref{lem:w}, these
contributions are bounded above by
\begin{equation}
    L^{j(-3-3+p+q)}O(L^{j(5-k)})
    = L^{j(k-5)}O(L^{j(5-k)})= O(1),
\end{equation}
which proves \refeq{kDpq}.
(In \refeq{LTwnstars}, the marginal term $w^{(5,**)}$ arises for
$\hat\kappa_\Delta^{33}$.)

\smallskip\noindent \emph{Proof of \refeq{k113}.}
The coefficient $\hat\kappa_1^{13}$ violates the assumption $p+q-i<3$
of \refeq{kipq}.
However, this coefficient is zero because
$F(\varphi_x^2,\varphi_y^6)$ and
$F(\varphi_x^6,\varphi_y^2)$ contribute no $\varphi^2$ term because the
maximal number of field derivatives in \refeq{Fexpand1Vpt} is two on each factor, and this leaves
$\varphi^4$, not $\varphi^2$.

\smallskip\noindent \emph{Proof of \refeq{kipD}.}
There are three cases:  $F(\tau_\Delta,\tau_\Delta)$,
$F(\tau_\Delta, \varphi^{2q})$, and $F(\varphi^{2p},\tau_\Delta)$.
For the first,
\begin{align}
    \LT_x \sum_y F(\Delta \varphi_x^2, \Delta \varphi_y ^2)
    & =  \LT_x  \sum_y \Delta_x\Delta_y  F(\varphi_x^2,\varphi_y^2 )
    \nnb
    & = \sum_y \Delta_x\Delta_y \LT_x  \left( c_1 \varphi_x  w_{xy} \varphi_y
    +
    c_2 w_{xy}^2
    \right)
    \nnb & =
    \Delta_x \sum_y \Delta_y w_{xy} (c_1 \varphi_x^2 +c_1' (y-x)^2\varphi_{\Delta,x}^2) +
    \Delta_x \sum_y \Delta_yc_2 w_{xy}^2
    \nnb & = 0,
\end{align}
with the last equality a consequence of summation by parts.
For the second case,
\begin{align}
    \LT_x \sum_y F(\Delta_x \varphi_x^2,  \varphi_y ^{2q})
    & =  \LT_x  \sum_y \Delta_x  F(\varphi_x^2,\varphi_y^{2q} )
    \nnb
    & = \sum_y \Delta_x \LT_x \left(c_1 \varphi_x  w_{xy} \varphi_y^{2q-1}
    +
    c_2 w_{xy}^2\varphi_y^{2q-2}
    \right).
\end{align}
If $q=3$ then $\LT_x$ simply replaces $y$ by $x$ and the result is zero
because $\sum_y \Delta_x w_{xy}^m=\sum_y \Delta_y w_{xy}^m=0$.
If $q=2$ then $\LT_x$ replaces  $\varphi_y^3$
by $\varphi_x^3$ in the first term, and replaces $\varphi_y^2$ by
by $\varphi_x^2 + (y-x)^2\Delta \varphi_x^2$ in the second term.
The overall result is again zero.
If $q=1$ then the $w^2$ term vanishes due to the Laplacian,
and the other term becomes
$\varphi_x^2 (\Delta w)^{(1)} + \varphi_\Delta^2 (\Delta w)^{(1,**)}$.
The first of these terms is zero, and the second yields a coefficient
$\hat\kappa_\Delta^{\Delta,2}$ which by Lemma~\ref{lem:w} is at most
\begin{equation}
  L^{-j(3+3-3-1)}(\Delta w)^{(1,**)} =  L^{-j(3+3-3-1)}O( L^{j2}) = O(1) .
\end{equation}
Finally, for the third case,
\begin{equation}
    \LT_x \sum_y F( \varphi_x^{2p}, \Delta\varphi_y ^{2})
    = \sum_y \Delta_y \LT_x \left( c_1 \varphi_x^{2p-1}  w_{xy} \varphi_y
    +
    c_2 \varphi_x^{2p-2} w_{xy}^2
    \right).
\end{equation}
The last term vanishes due to the Laplacian.  For $p \ge 2$ the effect of $\LT_x$
on the first term is the replacement of $\varphi_y$ by $\varphi_x$
and the result vanishes due to the Laplacian.  For $p=1$ the first term is
\begin{equation}
    \sum_y \Delta_y w_{xy}(c_1 \varphi_x^2 + c_1' (y-x)^2 \varphi_\Delta^2)
\end{equation}
and again this vanishes by summation by parts.
\end{proof}

\section{Analysis of the RG flow}
\label{sec:flow-analysis}

In this section, we analyse the RG flow in preparation for the proof of
Theorems~\ref{thm:main-n}--\ref{thm:tricrit-pt}.
The bulk flow is discussed in Section~\ref{sec:barflow}, and the observable
flow in Section~\ref{sec:flowobs}.

\subsection{Analysis of bulk flow}
\label{sec:barflow}

In this section, we analyse the approximate flow of Proposition~\ref{prop:barflow},
modified by inclusion of remainder terms produced by Theorem~\ref{thm:VK-bulk}.
We write the variables in the modified approximate flow
as $\mg$ rather than $\bar\gamma$.
By Proposition~\ref{prop:barflow}, \refeq{p2def} and Theorem~\ref{thm:VK-bulk},
the modified approximate flow is
\begin{align}
\lbeq{abarflow2}
    \mg_{3,+} & = \mg_3 - \bar\beta_3^{33} \mg_{3}^2 +e_3,
    \\
\lbeq{zbarflow2}
    \mg_{\Delta,+} & = \mg_{\Delta}
    - \bar\beta_\Delta^{33}\mg_{3}^2 + e_\Delta,
    \\
\lbeq{gbarflow2}
    \mg_{2,+} & =
    L\left( \mg_2(1 - p_2 \bar\beta_3^{33}\mg_3)
    + \bar\beta_2^3 \mg_3
    - \bar\beta_2^{33}\mg_3^2 \right) + e_2,
    \\
\lbeq{nubarflow2}
    \mg_{1,+} & =
    L^2 \left( \mg_{1}
    - \rho_1
    \right) ,
\end{align}
with
\begin{equation}
    \rho_1 =
    -\bar\beta_1^2 \mg_2
    -\bar\beta_1^3 \mg_3
    +\bar\beta_1^{22}\mg_2^2 +
    \bar\beta_1^{23}\mg_2\mg_{3}
    +
    \bar\beta_1^{33}\mg_{3}^2 - e_1,
\end{equation}
with all coefficients $\bar\beta = O(\chicCov)$,
and with $e_*=O(\vartheta\mu_3^3)$.  The approximate flow is the
special case with $e_*=0$.

A solution to the $\mg_3$ flow is given in \cite[Section~6.1.1]{BBS-brief},
which in our present setting yields the statements in the following proposition.
Recall from \refeq{b333def} that $\bar\beta_3^{33} = {\sf b} \delta[w^{(3)}]$
with ${\sf b}=18n+132$.
The $L$-dependent constant $\lambda_3$
in Proposition~\ref{prop:gjtj}(i) arises in Lemma~\ref{lem:w3}.

\begin{prop} \label{prop:gjtj}
Let $m^2 \in [0,\delta)$ and $\mg_{3,0}\in (0,\delta)$.

\smallskip \noindent
(i)
If $m^2=0$ then $\mg_{3,j}(0) \sim 1/({\sf b}\lambda_3 j) \to 0$
as $j \to \infty$.
For $m^2 >0$, the limit $\mg_{3,\infty}(m^2) = \lim_{j\to\infty}
\mg_{3,j}(m^2) > 0$ exists, is continuous in $m^2$,
 and obeys $\mg_{3,\infty}(m^2) \sim 1/({\sf b}w^{(3)}_\infty(m^2))$
as $m^2 \downarrow 0$.

\smallskip \noindent (ii)
The sequence $\mg_{3,j}$ obeys $\mg_{3,j} = O(\mg_{3,0})$,
$\mg_{3,j+1} = \mg_{3,j}(1+O(\mg_{3,0}))$,
$\vartheta_j(m^2) \mg_{3,j}(m^2) \le O(\mg_{3,j}(0))$, and
  \begin{align} \label{e:gbarsum}
    \sum_{l=j}^{\infty} \vartheta_l \mg_{3,l}^p
    &\leq O(\vartheta_j \mg_{3,j}^{p-1}) \qquad (p>1)
    .
  \end{align}
\end{prop}

\begin{lemma}
\label{lem:a0}
If $|f_j|\le O(A^{-j})$ for some $A>1$, and if $p>0$, then
\begin{equation}
    \sum_{j=0}^\infty f_j \mg_{3,j}^p = \mg_{3,0}^p\sum_{j=0}^\infty f_j +O(\mg_{3,0}^{p+1}).
\end{equation}
\end{lemma}

\begin{proof}
We apply \cite[(13.6.12)]{BBS-brief}
with $\psi(t) = p t^{p-1}$,
and then use \refeq{gbarsum} and the fact that
$\beta_{3,l}^{33} = O(\chicCov_l)$ by Proposition~\ref{prop:barflow}, to obtain
\begin{align}
    \mg_{3,0}^p - \mg_{3,j}^p
    & =
    p
    \sum_{l=0}^{j-1} (\beta_{3,l}^{33}\mg_{3,l}^2-e_{3,l})\mg_{3,l}^{p-1}
    + O(\mg_{3,0}^{p+1})
    \nnb & =
    p
    \sum_{l=0}^{j-1} \beta_{3,l}^{33}  \mg_{3,l}^{p+1}
    + O(\mg_{3,0}^{p+1}).
\end{align}
Therefore,  after interchanging sums and using $\mg_{3,l} = O(\mg_{3,0})$,
we obtain
\begin{align}
    \sum_{j=0}^\infty f_j (\mg_{3,0}^p - \mg_{3,j}^p)
    & =
    \sum_{l=0}^\infty \beta_{3,l}^{33}  \mg_{3,l}^{p+1}
    O(f_l)
    + O(\mg_{3,0}^{p+1}) = O(\mg_{3,0}^{p+1}),
\end{align}
as required.
\end{proof}

The other three equations can be solved backwards with zero final condition.
For $\mg_\Delta$, this gives
\begin{equation}
  \mg_{\Delta,j}
  =
  \sum_{l=j}^{\infty} (\bar\beta_{\Delta,l}^{33}\mg_{3,l}^2- e_{\Delta,l})
  ,
\end{equation}
which converges by \refeq{gbarsum}.
For $\mg_2$, we write the equation backwards and solve with zero final condition to get
\begin{align}
\lbeq{G2j}
    \mg_{2,j} & =
    \sum_{l=j}^{\infty} L^{-(l-j)}\pi_{j,l}^{-1}
    (- \bar\beta_{2,l}^3\mg_{3,l}+ \bar\beta_{2,l}^{33}\mg_{3,l}^2 -e_{2,l})
\end{align}
with
\begin{equation}
  \pi_{i,j}
  = \prod_{k=i}^j (1-p_2 \bar\beta_{3,k}^{33} \mg_{3,k})
  .
\end{equation}
By \cite[Lemma~6.1.6]{BBS-brief},
\begin{equation}
\lbeq{p2prod}
  \pi_{i,j}
  = \left( \frac{\mg_{3,j+1}}{\mg_{3,i}} \right)^{p_2}
  (c_i + O(\chicCov_j\mg_{3,j}))
  \quad \text{with} \quad
  c_i=1+O(\chicCov_i \mg_{3,i}).
\end{equation}
Finally,
\begin{align}
\lbeq{G1j}
    \mg_{1,j} & =
    \sum_{l=j}^\infty L^{-2(l-j)}
    \rho_{1,l}.
\end{align}
The powers of $L$ give exponential convergence of the sums in \refeq{G2j} and \refeq{G1j}.

\subsection{Analysis of observable flow}
\label{sec:flowobs}

The bulk flow has been constructed above, with
the critical initial conditions given by Theorem~\ref{thm:VK-bulk}.
We now construct the observable flow in terms of the bulk flow.

\subsubsection{Flow of $\lambda$}

The perturbative flow of $\lambda$ is as in \cite[Proposition~3.2]{ST-phi4}
(or \cite[(3.34)]{BBS-rg-pt} for $n=0$).  Namely, for $x=\pp,\qq$,
\begin{align}
  \label{e:lampt}
  \lambda_{x,\pt}
  &
  =
  \begin{cases}
    (1-\delta[\nu w^{(1)}]) \lambda_{x}  & (j+1 < j_{\pp\qq})
    \\
    \lambda & (j+1 \ge j_{\pp\qq}).
  \end{cases}
\end{align}
Here $j$ refers to the scale of the input $V$,
$w^{(1)}$ is given by \refeq{wndef} and is $O(L^{2j})$ by Lemma~\ref{lem:w}, and
\begin{equation}
    \delta_j[\nu w^{(1)}]
    =
    \nu_j^+ w_{j+1}^{(1)}
    -
    \nu_j w_{j}^{(1)}
\end{equation}
with $\nu_j^+$ defined to be the first order part of \refeq{gbarflow2}, namely
\begin{equation}
    \nu_j^+=\nu_j +
    \bar\beta_{1,j}^2 L^{-j}g_j  + \bar\beta_{1,j}^3 L^{-2j}\csix_j .
\end{equation}
Here $\nu_j^+$ contains an $a$-dependent term which is absent
absent in \cite[(3.24)]{BBS-rg-pt} where there was no $|\varphi|^6$ term.

Note that the perturbative flow of $\lambda$ stops one scale prior to
the coalescence scale.
Since $\pp,\qq$ are points in the torus $\Lambda_N$, we always have
$N \ge j_{\pp\qq}$, so the perturbative flow stops before
reaching scale $N$.
By Theorem~\ref{thm:VK-obs}, the nonperturbative flow also stops
prior to the coalescence scale and is given by
\begin{align}
  \label{e:lamptr}
  \lambda_{x,j+1}
  &
  =
    (1 - \delta_j[\nu w^{(1)}])\lambda_{x,j} + R^{\lambda_x}_{j+1}  \qquad (j+1 < j_{\pp\qq}),
\end{align}
with, for some $M=M(L)$,
\begin{align}
\lbeq{RlamM}
    |R^{\lambda_x}_{j+1}| & \le M\chicCov_j a_j^2 \1_{j+1< j_{\pp\qq}} .
\end{align}

Let
\begin{equation}
    f_j = 1 - \delta_j[\nu w^{(1)}], \qquad \Pi_{j} = \prod_{k=0}^jf_k.
\end{equation}

\begin{lemma}
\label{lem:lamprod}
There exists $\alpha = 1+O(a_0)$, independent of $j$, such that
\begin{equation}
    \Pi_{j} = \alpha (1+O(\chicCov_j a_j)).
\end{equation}
\end{lemma}

\begin{proof}
Let $\delta_j = \delta_j[\nu w^{(1)}]$,
$\delta_j' = \nu_{j+1}w_{j+1}^{(1)} - \nu_j w_j^{(1)}$, and
$\Delta_j = (\delta_j'-\delta_j)(1-\delta_j')^{-1}$.
Then $\delta_j'=O(\chicCov_j a_j)$ by estimating term by term
using $\nu_j=O(L^{-2j} \chicCov_j a_j)$ and $w_j^{(1)}=O(L^{2j})$
 by Theorem~\ref{thm:VK-bulk} and Lemma~\ref{lem:w}.
Also, $\Delta_j = O(\chicCov_j a_j^2)$ because
$\delta_j'-\delta_j = (\nu_{j+1}-\nu_j^+)w_{j+1}^{(1)}$ and there is a cancellation
of all first order terms in $\nu_{j+1}-\nu_j^+$.

By definition,
\begin{align}
    \Pi_j & = \prod_{k=0}^j(1-\delta_k') \prod_{i=0}^j (1 +   \Delta_i)
    =
    \exp\left[ \sum_{k=0}^j \log(1-\delta_k') \right]
    \exp\left[ \sum_{i=0}^j \log(1 +  \Delta_i) \right].
\end{align}
We use a telescoping sum and $w_0^{(1)}=0$ to see that
\begin{align}
    \sum_{k=0}^j \log(1-\delta_k')
    = -\nu_{j+1}w_{j+1}^{(1)} + \sum_{k=0}^j [\log(1-\delta_k') +\delta_k']
    .
\end{align}
The sum on the right-hand side has terms $O(\chicCov_k a_k^2)$ so it is summable
by \refeq{gbarsum},
the infinite sum is a constant $c=O(a_0)$
and the sum over $k >j$ is
$O(\chicCov_j a_j)$. Thus, with $\alpha = e^c$, we have
\begin{align}
    \Pi_j & =e^{-\nu_{j+1}w_{j+1}^{(1)}} \alpha e^{O(\chicCov_j a_j)},
\end{align}
which has the desired form since $\nu_{j+1}w_{j+1}^{(1)}
=O(\chicCov_j a_j)$.
\end{proof}

We define $\lambda^*_{\pp,j}$ to be the sequence $\lambda_{\pp,j}$ when
$\lambda_{\pp,0}=1$, $\lambda_{\qq,0}=0$
and $j_{\pp\qq}=\infty$.  As in \cite{BBS-saw4,ST-phi4}, the sequence $\lambda_{\pp,j}$
with $\lambda_{\pp,0}=\lambda_{\qq,0}=1$ is equal to $\lambda^*_{\pp,j}$ until the coalescence
scale.

\begin{prop}
\label{prop:lambda1}
For $N \in \N$ and $j \le N$,
$\lambda_{\pp,j}^*$ is continuous in $(m^2,a_0) \in [0,\delta)^2$  and
\begin{equation}
    \lambda_{\pp,j}^* = 1 + O(\chicCov_j a_j),
\end{equation}
and similarly for $\lambda_{\qq,j}^*$.
In particular,
$\lambda_{x,j_{\pp\qq}}  = 1 + O(\chicCov_{j_{\pp\qq}} a_{j_{\pp\qq}})$
for $x=\pp,\qq$.
\end{prop}

\begin{proof}
We first use induction on $j$ to prove that
\begin{equation}
\lbeq{lamIH}
    \lambda^*_{x,j+1} =  \Pi_j \left(1 + \sum_{k=0}^j e_k \right),
    \qquad
    |e_k| \le 2M \chicCov_k a_k^2
    \qquad (j+1 \le N)
    .
\end{equation}
If \refeq{lamIH} holds
for $j$ (it clearly holds for $j=0$), then, by \refeq{lamptr},
\begin{equation}
\lbeq{lamind1}
    \lambda^*_{x,j+1}
    =
    (1-\delta_j) \Pi_{j-1} \left( 1 + \sum_{k=0}^{j} e_k\right)
\end{equation}
with
$e_j =  \Pi_j^{-1} R^{\lambda_x}_{j+1}$.
This completes the induction, since by \refeq{RlamM} and Lemma~\ref{lem:lamprod},
\begin{equation}
    |e_{j}| \le 2|R^{\lambda_x}_{j+1}| \le 2 M\chicCov_j a_j^2.
\end{equation}

By \refeq{lam1identity},
\begin{align}
    \hat\chi_N
    &=
    \frac{1}{m^2}
    \frac{\partial_{\sigma_\pp} D Z_N(0;\hat\1)}{Z_N^\varnothing(0)}
    .
\end{align}
As in the proof of Corollary~\ref{cor:suscept},
$\lim_{N\to\infty} Z_N^\varnothing(0)=1$.
The $\sigma_\pp$ term in $e^{-V_N(\varphi)}$ is simply
$\lambda_{\pp,N}^*\sigma_\pp(\onehat\cdot \varphi)$,
so its double derivative (with respect to $\sigma_\pp$ and with respect
to $\varphi$ in the direction $\hat\1$) is $\lambda_{\pp,N}^*$.
Since $\partial_{\sigma_\pp} D Z_N= \partial_{\sigma_\pp} D I_N
+\partial_{\sigma_\pp} D K_N$, we can use
Theorem~\ref{thm:VK-obs} to take the limit $N \to \infty$ and obtain
\begin{equation}
    \hat \chi =  m^{-2} \lambda_{\pp,\infty}^*.
\end{equation}
Thus, from \refeq{chi-m} we conclude that $\lambda_{\pp,\infty}^*=1$.
By \refeq{lamIH}, Lemma~\ref{lem:lamprod}, and \refeq{gbarsum}, this implies that
\begin{align}
    \lambda_{x,j+1}^* - 1
    & =
    \alpha(1+O(\chicCov_ja_j))\left(1 + \sum_{k=0}^j e_k \right)
    -
    \alpha \left(1 + \sum_{k=0}^\infty e_k \right)
    \nnb & =
    -
    \alpha \sum_{k=j+1}^\infty e_k  +O(\chicCov_ja_j)
     = O(\chicCov_ja_j).
\end{align}

The remaining item is the continuity.  The continuity can be concluded along the lines
of the corresponding argument in the proof of \cite[Proposition~4.3]{BBS-saw4},
and we omit the details.
\end{proof}

\subsubsection{Flow of $q$}

The perturbative flow of $q$ is exactly as in \cite[Proposition~3.2]{ST-phi4}
(or \cite[(3.35)]{BBS-rg-pt} for $n=0$), namely, for $x=\pp,\qq$,
\begin{align}
  \label{e:qpt}
  q_{x,\pt}
  &
  =
  q_{x}
  +
  \lambda_\pp \lambda_\qq
  \, C_{j+1;\pp,\qq}
,
\end{align}
with $j$ the scale of the input $V$.
Since $C_{j+1;a,b}=0$ when $j+1 \le j_{\pp\qq}$,
for $j \le j_{\pp\qq}$ we have $q_{\pt} = 0$ if $q=0$.
Thus, while the perturbative flow of $\lambda$ stops at the coalescence scale,
the flow of $q$ only starts at the coalescence scale.
By Theorem~\ref{thm:VK-obs}, the nonperturbative flow also starts
at the coalescence scale, and
\begin{align}
  \lbeq{qptr}
  q_{x,j+1} & = q_{x,j} + \lambda_{\pp,j_{\pp\qq}} \lambda_{\qq,j_{\pp\qq}}
  \, C_{j+1;\pp,\qq}
  + R^{q_x}_{j+1} \qquad (j+1 \ge j_{\pp\qq}),
\end{align}
with
\begin{align}
\lbeq{RqbdM}
    |R^{ q_x}_{j+1}| & \le
    \1_{j+1\ge j_{\pp\qq}}
    |\pp-\qq|^{-1} 4^{-(j-j_{\pp\qq})} O(\chicCov_j a_j)
     .
\end{align}

\begin{prop}
\label{prop:q}
Let $\lambda_{\pp,0}=\lambda_{\qq,0}=1$.
For $(m^2,a_0) \in [0,\delta)^2$ and $x=\pp,\qq$, the limit
\begin{align}
    q_{x,\infty}(m^2,a_0)  &= \lim_{N \to \infty}q_{x,N}(m^2,a_0),
\end{align}
exists, is continuous, and, as $|\pp-\qq|\to\infty$,
\begin{equation}
\lbeq{qLap}
  q_{x,\infty}(0, a_0)
  =
  (-\Delta_{\Z^3}^{-1})_{\pp,\qq}
  \left( 1 + O \left( \frac{1}{\log |\pp-\qq|} \right) \right)
  .
\end{equation}
\end{prop}

\begin{proof}
For $N \in \N$ and  $(m^2,g_0) \in [L^{-2(N-1)},\delta) \times (0,\delta)$,
the solution of the recursion \refeq{qptr} is
\begin{equation}
    q_{x,N}
    =
    \lambda_{\pp, j_{\pp\qq}} \lambda_{\qq, j_{\pp\qq}}  w_{N; \pp,\qq}
    + \sum_{i = j_{\pp\qq}}^{N - 1} R^{q_x}_{i+1}.
\end{equation}
The limit $N \to\infty$ exists, and
by Proposition~\ref{prop:lambda1}, \refeq{RqbdM}, and Proposition~\ref{prop:gjtj},
\begin{align}
    q_{x,\infty}(m^2,a_0)
    &=
    \lambda_{\pp, j_{\pp\qq}} \lambda_{\qq, j_{\pp\qq}}  C_{\pp,\qq}(m^2)
    + \sum_{i = j_{\pp\qq}}^{\infty} R^{q_x}_{i+1}
    \nnb
    & =
    (1+O(\chicCov_{j_{\pp\qq}}a_{j_{\pp\qq}}))C_{\pp,\qq}(m^2)
    +
    |\pp-\qq|^{-1}O(\chicCov_{j_{\pp\qq}}a_{j_{\pp\qq}})
    \nnb
    & =
    (1+O(j_{\pp\qq}^{-1}))C_{\pp,\qq}(m^2)
    +
    |\pp-\qq|^{-1}O(j_{\pp\qq}^{-1})
    .
\end{align}
The continuity is a consequence of the continuity of
$C_{\pp,\qq},\lambda_{x,\pp\qq}$ and $R^{q_x}$, and
the estimate \refeq{qLap} follows from $j_{\pp\qq} \asymp_L \log|\pp-\qq|$ and
the fact that
$C_{\pp,\qq}(0) = (-\Delta_{\Z^3})^{-1}_{\pp,\qq}$.
\end{proof}

\section{Proof of Theorems~\ref{thm:main-n}--\ref{thm:tricrit-pt}}

\subsection{Proof of Theorem~\ref{thm:main-n}}
\label{sec:pf-main-n}

Fix $(m^2,a) \in (0,\delta)$.
Let $(a_0,g_0,\nu_0,z_0)$ be given by the continuous functions
$(a_0^*,g_0^*,\nu_0^*,z_0^*)$ of Proposition~\ref{prop:changevariables-i};
we record this with notational stars in the following.
By \refeq{corrdiff}, and since $W_N$ has no $\sigma_\pp  \sigma_\qq$ term by
\cite[Proposition~4.10]{BS-rg-IE},
\begin{align}
    G_{N;\pp,\qq}(a,g^*,\nu^*;n)
    &=
    (1+z_0^*) 
    \partial_{\sigma_\pp\sigma_\qq}^2 \log Z_N^*(0)
    \nnb & =
    (1+z_0^*) 
    \partial_{\sigma_\pp\sigma_\qq}^2 \log
    \left(
    e^{\frac{1}{2}(q_{\pp,N}^*+q_{\qq,N}^*)\sigma_\pp\sigma_\qq}(1+W_N^*(0)) +K_N^*(0)
    \right)
    \nnb & =
    (1+z_0^*)
    \left( \frac{1}{2}(q_{\pp,N}^*+q_{\qq,N}^*) +
     \partial_{\sigma_\pp\sigma_\qq}^2 \log(1 +K_N^*(0)) \right)
    .
\end{align}
By Proposition~\ref{prop:q} and Theorem~\ref{thm:VK-obs},
the $N \to \infty$ limit of the right-hand side exists and is
\begin{align}
    G_{\pp,\qq}(a,g^*,\nu^*;n)
    &=
    (1+z_0^*)
    \frac{1}{2}(q_{\pp,\infty}^*+q_{\qq,\infty}^*)
    .
\end{align}
Since $z_0^*$ and $q_{x,\infty}^*$ are continuous
as $m^2 \downarrow 0$, from \refeq{qLap} we obtain
\begin{align}
    G_{\pp,\qq}(a,g^*(0,a),\nu^*(0,a);n)
    &=
    (1+z_0^*(0,a))
    \frac{1}{2}(q_{\pp,\infty}(0,a_0^*(0,a))+q_{\qq,\infty}(0,a_0^*(0,a)))
    \nnb & =
    (1+z_0^*(0,a))
    (-\Delta_{\Z^3}^{-1})_{\pp,\qq}
    \left( 1 + O \left( \frac{1}{\log |\pp-\qq|} \right) \right)
    .
\end{align}
This proves \refeq{G-asy} with $A_{a,n}=(1+z^*(0,a))(4\pi)^{-1}$, since
$(-\Delta_{\Z^3})^{-1}_{\pp,\qq} = \frac{1}{4\pi}
|\pp-\qq|^{-1}(1+O(|\pp-\qq|^{-2}))$ (see, e.g., \cite{Lawl91}, the
constant $(4\pi)^{-1}$ is for our definition of the Laplacian).
The tricritical point is $(g^*(0,a),\nu^*(0,a))$, and the
continuous curve in Theorem~\ref{thm:main-n} is the curve
$(g^*(m^2,a),\nu^*(m^2,a))$ parametrised by $m^2$.

\subsection{Proof of Theorem~\ref{thm:tricrit-pt}}
\label{sec:tricrit-point}

Throughout this section we take $m^2=0$,
and we write $\simeq$ for equality up to an additive term that is $O(\mg_{3,0}^2)$.

It suffices to prove that
\begin{align}
\lbeq{g0asy}
    \mg_{2,0} & \simeq  -\frac{3}{2} (n+4)C_{0,0}\mg_{3,0},
    \\
\lbeq{nu0asy}
    \mg_{1,0} &\simeq \frac 34 (n+4)(n+2) C_{0,0}^2 \mg_{3,0}.
\end{align}
The proof is similar to the analysis in \cite[Section~8.5]{BBS-saw4-log}.
Our starting point is the equations
\begin{align}
\lbeq{G20}
    \mg_{2,j} & =
    -
    \sum_{l=j}^{\infty} L^{-(l-j)}\pi_{j,l}^{-1}
    (\bar\beta_{2,l}^3\mg_{3,l}
    +
    O(\chicCov_l\mg_{3,l}^2)
    )
    ,
\\
\lbeq{G10}
    \mg_{1,0} & =
    -
    \sum_{l=0}^\infty L^{-2l}
    (\bar\beta_{1,l}^2 \mg_2
    +\bar\beta_{1,l}^3 \mg_3 + O(\chicCov_l\mg_{3,l}^2)
    )
    ,
\end{align}
which are consequences of \refeq{G2j} and \refeq{G1j}.
Here,
\begin{equation}
\lbeq{p2prod0}
    \pi_{j,l}^{-1}
  = \left( \frac{\mg_{3,j}}{\mg_{3,l+1}} \right)^{p_2}
  (c_j^{-1} + O(\chicCov_l\mg_{3,l}))
  \quad \text{with} \quad
  c_j^{-1}=1+O(\chicCov_j \mg_{3,j}) ,
\end{equation}
and
\begin{align}
    L^{-2l} \bar\beta_{1,l}^2 & = b_1^2 C_{l+1} ,
    \quad
    L^{-2l} \bar\beta_{1,l}^3  = b_1^3 C_{l+1}^2,
    \quad
    L^{-l}\bar\beta_{2,l}^3  = b_2^3 C_{l+1}
\end{align}
with   $C_{l+1}=C_{l+1;0,0}$ and
\begin{align}
    b_1^2 & = (n+2),
    \quad
    b_1^3  = \frac 34 (n+4)(n+2),
    \quad
    b_2^3  = \frac{3}{2} (n+4).
\end{align}
Note that $b_1^2b_2^3 = 2b_1^3$.

\begin{proof}[Proof of Theorem~\ref{thm:tricrit-pt}]
By \refeq{G20}, \refeq{p2prod}, and Lemma~\ref{lem:a0},
\begin{align}
    \mg_{2,0} & \simeq -b_2^3 \sum_{j=0}^\infty \pi_{0,j}^{-1}C_{j+1}\mg_{3,j}
    \nnb & =
    -b_2^3(1+O(\mg_{3,0})) \mg_{3,0}^{p_2} \sum_{j=0}^\infty C_{j+1}\mg_{3,j}^{1-p_2}
    \simeq
    -b_2^3  \mg_{3,0}   C_{0,0}
     ,
\end{align}
which proves \refeq{g0asy}.
Similarly, by \refeq{G20}--\refeq{G10}
(we use $b_1^2b_2^3 = 2b_1^3$ to obtain the third line),
\begin{align}
    \mg_{1,0} & \simeq
    - b_1^2 \sum_{j=0}^\infty C_{j+1} \mg_{2,j} -  b_1^3 \sum_{j=0}^\infty C_{j+1}^2 \mg_{3,j}
    \nnb & \simeq
      b_1^2 b_2^3 \sum_{j=0}^\infty C_{j+1} \sum_{l=j}^\infty \pi_{l,j}^{-1}C_{l+1}  \mg_{3,l}
      -  b_1^3 \sum_{j=0}^\infty C_{j+1}^2 \mg_{3,j}
     \nnb & \simeq
      b_1^3
      \left(
      \sum_{j=0}^\infty 2C_{j+1} \sum_{l=j+1}^\infty \pi_{l,j}^{-1} C_{l+1} \mg_{3,l}
      + \sum_{j=0}^\infty C_{j+1}^2 \mg_{3,j}
      \right)
      \nnb &=
      b_1^3
      \sum_{l=0}^\infty \left(2C_{l+1} \sum_{j=0}^{l-1} \pi_{j,l}^{-1}  C_{j+1}  +  C_{l+1}^2
      \right)\mg_{3,l}
      .
\end{align}
By Lemma~\ref{lem:a0},
\begin{equation}
      \sum_{l=0}^\infty   C_{l+1}^2 \mg_{3,l}
      \simeq
      \mg_{3,0}\sum_{l=0}^\infty   C_{l+1}^2.
\end{equation}
Also,
\begin{align}
     2\sum_{l=0}^\infty C_{l+1} \sum_{j=0}^{l-1} \pi_{j,l}^{-1}  C_{j+1}  \mg_{3,l}
     & =
     (2+O(\mg_{3,0}))
     \sum_{l=0}^\infty C_{l+1} \mg_{3,j}^{p_2}\sum_{j=0}^{l-1} C_{j+1}  \mg_{3,l}^{1-p_2}
     \nnb & =
     (2+O(\mg_{3,0}))
     \sum_{l=0}^\infty C_{l+1} \mg_{3,j}^{p_2}  \left(\sum_{j=0}^{l-1} C_{j+1}
     \mg_{3,0}^{1-p_2} +O(\mg_{3,0}^{2-p_2}) \right)
     \nnb & \simeq
     2\mg_{3,0} \sum_{l=0}^\infty C_{l+1} \sum_{j=0}^{l-1} C_{j+1}
     =
     \mg_{3,0} \left( C_{0,0}^2 - \sum_{l=0}^\infty C_{l+1}^2 \right).
\end{align}
Therefore, there is a cancellation and
\begin{align}
    \mg_{1,0} &\simeq b_1^3 C_{0,0}^2 \mg_{3,0},
\end{align}
which is \refeq{nu0asy}.  This completes the proof.
\end{proof}

\appendix
\normalsize

\section{Covariance decomposition}
\label{app:decomp}

Finite-range covariance decompositions of the type we use were developed
in \cite{BGM04}, and a different perspective was given in
\cite{Baue13a} which we follow here.
The following lemma is a restatement of \cite[Proposition~6.1]{BBS-rg-pt}.
The $C_j$ are positive-definite on $\Zd$, as discussed below \cite[(1.7)]{MS00}.
This does not immediately imply that they are positive-definite on the torus, but they are
at least positive semi-definite and this is sufficient for our needs.

Let $\nabla_{x_k}$ denote the
finite-difference operator $\nabla_{x_k}f(x,y)=f(x+e_k,y)-f(x,y)$, where $e_k$ is
one of the $2d$ (positive or negative) unit vectors on $\Z^d$.
We write $\nabla_x^\alpha=\nabla_{x_1}^{\alpha_1} \dotsb \nabla_{x_d}^{\alpha_d}$ for a
multi-index $\alpha=(\alpha_1,\dotsc,\alpha_d)\in \N_0^d$.

\begin{prop}
\label{prop:Cdecomp}
  Let $d > 2$, $L\geq 2$, $j \ge 1$, $m^2 \in [0,1]$.
  There exist positive-definite covariances $C_j$
  on $\Zd$ such that $(-\Delta +m^2)^{-1} = \sum_{j=1}^\infty C_j$, and the following hold.
  \begin{enumerate}[(i)]
  \item
  For multi-indices $\alpha,\beta$ with
  $\ell^1$ norms $|\alpha|_1,|\beta|_1$ at most
  some fixed value $p$,
  and for any $q>0$,
  \begin{equation}
    \label{e:scaling-estimate}
    |\nabla_x^\alpha \nabla_y^\beta C_{j;x,y}|
    \leq c(1+m^2L^{2(j-1)})^{-q}
    L^{-(j-1)(d-2+(|\alpha|_1+|\beta|_1))},
  \end{equation}
  where $c=c(p,q)$ is independent of $m^2,j,L$.
  If $m^2L^{2(N-1)} \ge \varepsilon$ for some $\varepsilon >0$,
  then the same bound holds for $C_{N,N}$
  with $c$ depending on $\varepsilon$ but independent of $N$.

  \item
  The $C_j$ have the finite-range property $C_{j;x,y}=0$ if $|x-y| \ge \frac 12 L^j$.

  \item
    Let $m^2=0$.
    There exists a smooth function
    $c_0: \R^d \to \R$ with compact support such that, as $j\to\infty$,
    \begin{equation}
    \label{e:Cc}
      C_{j;x,y} = L^{-(d-2)j}c_0(L^{-j} (x-y)) + O(L^{-(d-1)(j-1)})
      \quad \text{for $m^2 =0$}.
    \end{equation}
  \end{enumerate}
\end{prop}

The next lemma concerns the quantities defined in \refeq{wndef}.
Recall that the mass scale $j_m$ is the largest integer $j$ such that $mL^{j} \le 1$,
with $j_0=\infty$.

\begin{lemma}
\label{lem:w}
For $d=3$, $j \ge 1$, and $m^2 \in [0,1]$,
\begin{alignat}{2}
    w_j^{(k)} &= O_L(  L^{(3-k)(j\wedge j_m)}) &\qquad& (k< 3),
    \\
    w_j^{(k,**)} &= O_L( L^{(5-k)(j\wedge j_m)}) &\qquad& (k< 5),
    \\
    (\Delta w_j)^{(k,**)} &= O_L(L^{(3-k)(j\wedge j_m)}) &\qquad& (k< 3).
  \end{alignat}
\end{lemma}

\begin{proof}
The factor involving the mass in \refeq{scaling-estimate} obeys
$(1+m^2L^{2(j-1)})^{-q} \le O_L(L^{-q(j-j_m)_+})$, so
$C_{j;0,x} = O(L^{-j}L^{-q(j-j_m)_+})$.
For $k=1$, with the finite-range property we obtain
\begin{equation}
    w_j^{(1)} = \sum_{1 \le i  \le j} \sum_x C_{i;0,x}
    \le
    \sum_{1 \le i  \le j} L^{3i} L^{-i}O_L(L^{-q(i-j_m)_+})
    \le O_L(L^{2(j\wedge j_m)}),
\end{equation}
and, for $k=2$,
\begin{align}
    w_j^{(2)} & \le 2 \sum_{1 \le i \le l  \le j} \sum_x C_{i;0,x} C_{l;0,x}
    \le
    \sum_{1 \le i \le l \le j} L^{3i} L^{-i}L^{-q(i-j_m)_+} L^{-l}O_L(L^{-q(l-j_m)_+})
    \nnb & \le
    \sum_{1 \le i  \le j} L^{2i}  L^{-q(i-j_m)_+} L^{-i}O_L(L^{-q(i-j_m)_+})
    \le O_L( L^{(j\wedge j_m)}).
\end{align}

The proof for $(\Delta w_j)^{(k,**)}$ is the same because $\sum_x C_{i;0,x}$ and
$\sum_x x_1^2 \Delta C_{i;0,x}$ obey the same bounds $L^{3i}L^{-i}=L^{2i}$ and
$L^{3i}L^{2i}L^{-3i}=L^{2i}$ (with the same enhanced decay beyond the mass scale).

The proof for $w_j^{(k,**)}$ is almost the same for $k=1,2$ as it is
for $w_j^{(n)}$: we simply replace the $L^{3i}L^{-i}$
bound on $\sum_x C_{i;0,x}$ by a bound with an additional factor $L^{2i}$.
For $k=3,4$ it is similar.
\end{proof}

With further effort, the constant $\lambda_3$ in Lemma~\ref{lem:w3} could be
computed as an explicit universal constant times $\log L$, as in
\cite[Lemma~6.3(a)]{BBS-rg-pt}.

\begin{lemma}
\label{lem:w3}
  Let $d=3$ and $m^2=0$.  There is an $L$-dependent constant $\lambda_3$ such that
  \begin{alignat}{2}
  \lbeq{w3}
    \lim_{j \to \infty} (w_{j+1}^{(3)} - w_j^{(3)}) &= \lambda_3. 
\end{alignat}
\end{lemma}

\begin{proof}
The proof
is similar to the proof of \cite[Lemma~6.3(a)]{BBS-rg-pt}
for $w^{(2)}$ when $d=4$, so we only give a sketch.
Constants in error estimates may depend on $L$ here, and we abbreviate
subscripts $0,x$ by $x$ alone.  Sums are over $x \in \Z^3$ and integrals are over $\R^3$.

Since $w_{j+1}=w_j+C_{j+1}$, we have
\begin{align}
    w_{j+1}^{(3)} - w_j^{(3)}
    & =
    \sum_x \left( C_{j+1;x}^3 + 3C_{j+1;x}^2 w_{j;x} + 3C_{j+1;x} w_{j;x}^2  \right)
    \nnb & =
    \sum_x C_{j+1;x}^3 + 3 \sum_{k=1}^j \sum_x C_{j+1;x}^2 C_{k;x}
    + 3 \sum_{k,l=1}^j \sum_x C_{j+1;x} C_{k;x} C_{l;x}.
\lbeq{dw3pf1}
\end{align}
We use \refeq{Cc} to write $C_{i;x}=c_i(x) + e_i(x)$ with
$c_i(x) = L^{-i}c_0(L^{-i}x)$ and $e_i(x)=O (L^{-2i})$.
By  Riemann summation,
\begin{align}
    \sum_x  c_k(x)c_{k+s}(x)c_{k+t}(x)
    & =
    L^{-3k} \sum_x  c_0(L^{-k}x)c_{s}(L^{-k}x)c_{t}(L^{-k}x)
    \nnb & =\int c_0(x)c_s(x)c_t(x) dx +O(L^{-k-s-t}),
\end{align}
where the error term is a bound on $L^{-k}\|c_0c_sc_t\|_\infty$.
This permits the sums over $x$ in \refeq{dw3pf1} to be approximated by integrals,
and the sums of those integrals over scales produce the constant $\lambda_3$.
For example, the first term on the right-hand side of \refeq{dw3pf1} is
\begin{align}
    \sum_x  C_{j+1;x}^3
    & =
    \sum_x c_{j+1}(x)^3
    +
    \sum_x e_{j+1}(x)\left( 3c_{j+1}^2(x) + 3 c_{j+1}(x)e_{j+1}(x) + e_{j+1}^2(x) \right).
\end{align}
The first term is equal to $\int c_0^3$ plus an error of order $L^{-j}$, and the
remaining sum is smaller by $L^{-j}$ because there is at least one $e_{j+1}$ factor.
Similarly, the middle term in \refeq{dw3pf1} is handled by
\begin{align}
    \sum_{k=1}^j \sum_x C_{j+1;x}^2 C_{k;x}
    & =
    \sum_{l=1}^j \int c_{l}^2 c_0 +O(L^{-j}),
\lbeq{dw3pf2}
\end{align}
together with the fact that the series $\sum_{l=1}^\infty \int c_{l}^2 c_0$
converges since its terms are $O(L^{-2l})$.
The last term in \refeq{dw3pf1} can be handled similarly.
\end{proof}

\section{Existence of a global RG flow}
\label{sec:step-flow}

We discuss the minor changes to
the analysis of \cite{BS-rg-IE,BS-rg-step,BBS-rg-flow,BBS-saw4}
that lead to a proof of Theorems~\ref{thm:VK-bulk}--\ref{thm:VK-obs}.

\subsection{Parameters}
\label{sec:circ}

Several parameters require natural adjustments
to take into account the change from $|\varphi|^4$ and $d=4$ to $|\varphi|^6$ and $d=3$.

\subsubsection{Norm parameters}
\label{sec:normparameters}

The space $\Kcal$ is as in \cite[Definition~4.5]{ST-phi4} with $h=\onehat$,
and $\CKspace$ is its restriction to connected polymers.
We modify the definition of
the $\Wcal$-norm on $\CKspace$, with the following
revised parameters compared to \cite[Section~1.7]{BS-rg-step}.
We fix  a sequence $\agen_j$ such that
\begin{equation}
\lbeq{agenbds}
    \frac 12 \agen_{j+1} \le \agen_j \le 2\agen_{j+1}.
\end{equation}
Specifically, given $\mgen^2 \ge 0$, we use the sequence
\begin{equation} \label{e:agendef}
  \agen_j(\mgen^2,\tilde\csix_0) =
  \bar{\csix}_j(0,\tilde\csix_0) \1_{j \le j_{\mgen}}
  + \bar{\csix}_{j_{\mgen}}(0,\tilde\csix_0) \1_{j > j_{\mgen}},
\end{equation}
where $j_m$ is the mass scale, and $\abar$ is the sequence $\mu_3$ of
\refeq{abarflow2} with remainder $e_3=0$ and initial condition $\abar_0=\tilde\csix_0$.
The norm depends on $\mgen^2$, which is a replacement for $m^2$ to avoid dependence of
the norm on $m^2$; see the discussion above \cite[Theorem~1.13]{BS-rg-step}.

Given a (large) $L$-dependent constant $\ell_0$ and
a (small) $L$-independent constant $k_0$, we set
\begin{align}
\lbeq{ellh-def}
    \h_j & =
    \begin{cases}
    \ell_0 L^{-j/2} & (\h=\ell)
    \\
    k_0 \agen_j^{-1/6} L^{-dj/6} & (\h=h),
    \end{cases}
\\
\label{e:hsigdef}
\h_{\sigma,j} &=
 \ell_{j \wedge j_{\pp\qq}}^{-1}
2^{(j - j_{\pp\qq})_+}
\times
\begin{cases}\agen_j & (\h = \ell)
\\
\agen_j^{1/6} & (\h = h).
\end{cases}
\end{align}
With $\thgen_j=2^{-(j-j_{\mgen})_+}$,
we define
\begin{equation}
\lbeq{epdVdef}
    \epdV_j =
    \begin{cases}
    \thgen_j \agen_j & (\h=\ell)
    \\
    \thgen_j \agen_j^{1/6} & (\h=h),
    \end{cases}
\end{equation}
and set
\begin{equation}
    \label{e:ratiodef}
    \omega_{j}
    = \frac{\epdV_{j}(\ell)}{\epdV_{j}(h)}
    = \agen_{j}^{5/6}.
\end{equation}

With the above parameters, exactly
as in \cite[Section~1.7]{BS-rg-step} we define a norm on $\Ccal\Kcal_j$ by
\begin{equation}
\label{e:9Kcalnorm}
    \|K\|_{\Wcal_{j}}
    =
    \max
    \Big\{
    \|K \|_{\Fcal_j(G)},\,
    \omega_{j}^{3}
    \|K \|_{\Fcal_j(\tilde{G})}
    \Big\}
.
\end{equation}
In \cite[(1.44)]{BS-rg-step} for $d=4$, \refeq{ratiodef} was
instead $\omega_j = \ggen_j^{3/4}$ so $\omega_j^3 = \ggen_j^{9/4}=O(\ell_j/h_j)^9$
and $\omega_j^3 (\ggen_j^{1/4})^3=\ggen_j^3$.
This determined that we needed $p_\Ncal \ge 9+1=10$ derivatives of the field
(see \cite[Lemma~2.4]{BS-rg-step}).
Now, instead, we have
$\omega_j^3$ of order $(\ell_j/h_j)^{15}$,
and we can choose any $p_\Ncal \ge 16$.

Also, a scale-dependent norm on $\Ucal \simeq \C^{10}$ is defined by
\begin{equation}
\lbeq{Vcalnormdef}
    \|U\|_\Ucal = \max\{ |\csix|,|z|,|y|, |g| L^{j}, |\nu|L^{2j}, |u|L^{3j} ,
    \ell_j\ell_{\sigma,j}|\lambda_\pp|,
    \ell_j\ell_{\sigma,j}|\lambda_\qq|,
     \ell_{\sigma,j}^2|q_\pp|  , \ell_{\sigma,j}^2|q_\qq| \},
\end{equation}
with $\ell_j$ and $\ell_{\sigma,j}$ as in \refeq{ellh-def}--\refeq{hsigdef}.
The RG domain for $V$ now becomes, with $C_\DV$ a (large) universal constant,
\begin{equation} \label{e:DVdef}
\begin{aligned}
  \DV_j
  &= \{(a,g,\nu,z,\lambda_a,\lambda_b) \in \C^6
  :
    {\rm Re}\, a >C_{\DV}^{-1} \agen , \; |{\rm Im}\, a| < \frac{1}{10} C_{\DV} \agen,
   \\
   & \qquad\qquad\qquad
   L^j |g|, L^{2j}|\nu|,|z| \le C_\DV \agen, \; |\lambda_a|,|\lambda_b| \le C_\DV
    \}
    .
\end{aligned}
\end{equation}
The stability domains \cite[(1.85)--(1.86)]{BS-rg-IE} are replaced now,
given $\alpha,\alpha',\alpha_L'' >0$, by
\begin{align}
\lbeq{DVell}
    \bar\DV_j(\ell) &=
    \{V  :
    \; |{\rm Im}\, \csix| < \textstyle{\frac {1}{5}} {\rm Re}\, \csix,
    \;
    \epsilon_{V,j}(\ell_j) \le \alpha_L'' \agen
    \},
\\
\lbeq{DVh}
    \bar\DV_j(h) &=
    \{V  :
    \; |{\rm Im}\, \tilde\csix| < \textstyle{\frac {1}{5}} {\rm Re}\, \tilde\csix,
    \;
    \alpha \le \epsilon_{\csix\tau^3,j}(h_j), \;
    \epsilon_{V,j}(h_j) \le \alpha'
    \}.
\end{align}

\subsubsection{Small parameters $\epV$ and $\epdV$}

The small parameter $\epV$, defined in \cite[(1.80)]{BS-rg-IE} as a sum of $T_0$-seminorms
of monomials, requires modification
in our present setting.  The parameters $\h$ and $\h_\sigma$ have been chosen
precisely to make this modification insignificant.  We illustrate this here
for the monomials $a\tau^3$, $g\tau^2$, and $\nu \tau$:
\begin{align}
    L^{dj} \left(
    \|\csix \tau^3\|_{T_0(\h)} + \|g \tau^2\|_{T_0(\h)} +  \|\nu \tau \|_{T_0(\h)}
    \right)
    & \asymp
    \begin{cases}
    |\csix| \ell_0^6 +  L^j |g| \ell_0^4 +   L^{2j} |\nu| \ell_0^2 & (\h=\ell)
    \\
    k_0^6  + L^j |g|a^{-4/6}  k_0^4 + L^{2j} |\nu| a^{-2/6} k_0^2 & (\h=h).
    \end{cases}
\end{align}
For $V\in \DV_j$,
the right-hand side is bounded by an $L$-dependent multiple of $\agen$ when $\h=\ell$,
and by an $L$-independent multiple of $k_0$ when $\h=h$.

Our choice of the small parameter $\epdV$ in \refeq{epdVdef} is made to dominate the norm
of $\delta V$ as in  \cite[Lemma~3.4]{BS-rg-IE}.
By following the proof of \cite[Lemma~3.4]{BS-rg-IE},
for $V \in \bar\DV$ we again get, as required,
\begin{align}
    \|\delta V(b) \|_{T_0(\h \sqcup \ell)}
    & \le \|\theta V(b) - V(b)\|_{T_0(\h \sqcup \ell)}
    + \|V(b)-\Vpt(b)\|_{T_0(\h \sqcup \ell)}
    \nnb  &
    \le
    \frac{\hat\ell}{\h} O_L(\epV) \le O_L(\epdV).
\end{align}

\subsubsection{Localisation parameters}
\label{sec:loc}

By definition, the operator $\Loc_X$ acts term by term in the direct sum decomposition
\refeq{Fdecomp}, with an action that depends on the number of $\sigma$ factors
(as well as on the scale when there is just one $\sigma$).
As discussed in detail in \cite{BS-rg-loc}, the definitions require:
(i) specification of the field dimensions,
(ii) choice of a maximal monomial dimension $d_+(\alpha)$ for each $\alpha
\in \{\varnothing, \pp,\qq,\pp\qq\}$, and
(iii) choice of covariant field polynomials $\hat P$.
Item~(iii) is done exactly as in \cite[(1.19)]{BS-rg-loc} and
plays a minor role.  The field dimension is always $\frac{d-2}{2}=\frac 12$ in this paper.
For (ii), we make the following choices.
For $\alpha=\varnothing$, we set $d_+(\varnothing) =d=3$.
For $\alpha = \pp\qq$, we set $d_+(\pp\qq) = 0$.
For $\alpha = \pp$ and $\alpha = \qq$,
we make the scale dependent choice
$d_+^{(j)}(\pp) = d_+^{(j)}(\qq) = \frac 12 \1_{j < j_{\pp\qq}}$,
where $j_{\pp\qq}$ is the coalescence scale.
This is as in \cite{BBS-saw4,ST-phi4}, after taking into account that the field
dimension here is $\frac 12$ rather than $1$.

\subsection{Stability}

The delicate stability estimate is \cite[Proposition~5.1(ii)]{BS-rg-IE},
which shows how the $|\varphi|^4$ term in $e^{-V}$ provides integrability for $d=4$.
The next proposition adapts its essential part to our present setting, in which
the $|\varphi|^6$ term stabilises the integral for $d=3$.  Only minor modifications
to the proof of \cite[Proposition~5.1(ii)]{BS-rg-IE} are needed.
In addition to the $T_\varphi$-seminorm, the statement involves the same $\Phi$
and $\tilde\Phi$ norms used in \cite{BS-rg-IE}.

\begin{prop}
Let $q \ge 0$, $B\in \Bcal_j$, $V \in \Ucal$.
Suppose that $|{\rm Im}\, a| \le \frac 12 {\rm Re}\, a$
and $\epV \le C \epsilon_{a\tau^3}$.  Then
\begin{equation}
    \|e^{-V(B)}\|_{T_\varphi}
    \le e^{O(1+q^3) \epV} e^{-q\epsilon_{a\tau^3}\|\varphi\|^2_{\Phi(B^\Box)}}
    e^{O(q)\epV \|\varphi\|^2_{\tilde\Phi(B^\Box)}}
    .
\end{equation}
\end{prop}

\begin{proof}
For notational simplicity, we present the proof for the case $n \ge 1$; a minor
adaptation applies for $n=0$.
Constants in this proof can depend on $n$.

We isolate the $|\varphi|^6$ term in $V$ as
$V = \frac 18 a |\varphi|^6 +Q$.
By the product property of the $T_\varphi$-seminorm,
\begin{equation}
    \|e^{-V(b)}\|_{T_{\varphi}}
    \le
    e^{\|Q(b)\|_{T_{\varphi}}}
    \prod_{x\in B}\|e^{-\frac 18 \csix \varphi^6_x}\|_{T_{\varphi}}
    .
\end{equation}
Let $\alpha = \frac 18 {\rm Re}\, a$, so $\frac 18 |a| \le \frac 32 \alpha$ by
hypothesis.  Let $t_x = |\varphi_x|/\h$, $H=(\h,\ldots,\h)$, and $P(t)=(t+\sqrt{n})^2$.
As in the proof of \cite[Proposition~3.9]{BS-rg-norm},
\begin{equation}
    \||\varphi_x|^2\|_{T_\varphi} = |\varphi_x+H|^2 \le \h^2 P(t_x)
\end{equation}
and
\begin{equation}
    \|e^{-\frac 18 \csix \varphi_x^6}\|_{T_{\varphi}}
    \le
    e^{-2\alpha|\varphi_x|^6} e^{\frac 32 \alpha \h^6 P(t_x)^3}
    =
    e^{\alpha \h^6 [-2t_x^6 + \frac 32 P(t_x)^3]}.
\end{equation}
Given any $q_1 \ge 0$, we can estimate the exponent on the right-hand side to obtain
\begin{equation}
    \|e^{-\frac 18 \csix \varphi_x^6}\|_{T_{\varphi}}
    \le
    e^{\alpha \h^6 [q_2 - q_1(t_x^2+t_x^4)]}
\end{equation}
for some $q_2=O(1+q_1^3)$.
We define
$\|\varphi\|_{L^p}^p = L^{-dj}\sum_{x\in B}(|\varphi_x|/\h)^p$.
Then, since $\epsilon_{a\tau^3}= \frac 18 |a|\h^6 L^{dj} \le \frac 32 \alpha \h^6 L^{dj}$,
\begin{equation}
    \prod_{x\in B}\|e^{-\frac 18 \csix \varphi^6_x}\|_{T_{\varphi}}
    \le
    e^{(O(1+q_1^3) -\frac 23 q_1(\|\varphi\|_{L^2}^2
    + \|\varphi\|_{L^4}^4))\epsilon_{a\tau^3}}.
\end{equation}

Since
for some $c>0$ we have
\begin{equation}
    \|Q(b)\|_{T_{\varphi}}
    \le
    c\epV (\|\varphi\|_{L^4}^{4}+1),
\end{equation}
and since $ \epV \le C\epsilon_{a\tau^3}$ by hypothesis, it follows that
\begin{equation}
\lbeq{lognormbd1}
    \log \|e^{-V(b)}\|_{T_{\varphi}}
    \le
    \left[
    O(1+q_1^3) -\frac 23 q_1 \|\varphi\|_{L^2}^2  -\frac 23 q_1 \|\varphi\|_{L^4}^4
    + cC  (\|\varphi\|_{L^4}^{4}+1)
    \right]
    \epsilon_{a\tau^3}
    .
\end{equation}
Now we choose $q_1=\frac 32 (q +cC)$  to obtain
\begin{equation}
\lbeq{lognormbd2}
    \log \|e^{-V(b)}\|_{T_{\varphi}}
    \le
    O(1+q^3)\epV -q \|\varphi\|_{L^2}^2  \epsilon_{a\tau^3}
    .
\end{equation}
Finally, as the proof of \cite[Proposition~5.1(ii)]{BS-rg-IE} we use
\begin{equation}
\label{e:Sob2}
    \|\varphi\|_{L^2}^2
    \geq
    \frac{1}{2c_2^2} \|\varphi\|_{\Phi(B^{\Box})}^2
    -
    \|\varphi\|_{\tilde \Phi(B^{\Box}) }^2
    ,
\end{equation}
and redefine $q$ to complete the proof.
\end{proof}

\subsection{Proof of Theorems~\ref{thm:VK-bulk}--\ref{thm:VK-obs}}

The proof proceeds in three steps which exactly parallel the analysis for $d=4$,
as follows.

\begin{enumerate}
\item
\emph{A single RG step.}  We estimate the map representing a single RG step
using the results of \cite[Section~1.8]{BS-rg-step}.  The domain for $V$
is given by \refeq{DVdef},
and $\ggen$ in \cite{BS-rg-step} becomes instead $\agen$.

\item\emph{Global bulk RG flow.}
The construction of the $(m^2,a_0)$-dependent
critical initial condition $(g_0^c,\nu_0^c,z_0^c)$ for the global bulk RG flow
(without observables) in Theorem~\ref{thm:VK-bulk}
is achieved using an adaptation of \cite[Theorem~1.4]{BBS-rg-flow}.
The adaptation of \cite{BBS-rg-flow} is
notational only, to take into account that there now are two relevant variables $g,\nu$
rather than just $\nu$.

\item
\emph{Global observable flow.}
The bulk flow is independent of the flow of the observable coupling constants $(\lambda,q)$.
Unlike the construction of the critical initial condition in Step~2, which involves
what is essentially a delicate implicit function theorem (couched in the context of
local existence theory for ODEs in \cite{BBS-rg-flow}),
the observable flow is simply solved forward
recursively from the initial condition.  This requires a relatively straightforward
induction argument, which can be carried out just as in \cite[Section~4]{BBS-saw4}
(or as in \cite{ST-phi4}).  In Theorem~\ref{thm:VK-obs},
we have isolated the estimates produced by the induction argument.  The part of
that argument involving the coupling constants is
given in Section~\ref{sec:flowobs} to illustrate the calculations that
lead to our main result.
\end{enumerate}

\section*{Acknowledgements}
The work of ML and GS was supported in part by NSERC of Canada.
The work of GS was partially supported by a grant from the Simons Foundation.
The authors would like to thank the Isaac Newton Institute for Mathematical
Sciences for support and hospitality during the programme ``Scaling limits,
rough paths, quantum field theory'' when work on this paper was
undertaken; this work was supported by EPSRC Grant Number EP/R014604/1.
GS is grateful to Takashi Kumagai and Ryoki Fukushima
for support and hospitality at the Research
Institute for Mathematical Sciences, Kyoto University,
where part of this work was carried out.
We thank David Brydges for helpful comments on a preliminary version
of the paper.


\end{document}